\documentclass[aps,twocolumn,superscriptaddress]{revtex4-2}

\usepackage{makecell} 
\usepackage{graphicx}
\usepackage{dcolumn}
\usepackage{bm}
\usepackage{caption} 
\usepackage{mathtools} 
\usepackage{tensor} 
\usepackage{amssymb, mathtools, mathrsfs}
\usepackage{amsmath} 
\usepackage{xcolor}
\usepackage{array} 
\usepackage{booktabs}
\usepackage{subcaption}
\usepackage[inline]{enumitem}
\usepackage{soul}
\usepackage[toc,page]{appendix}
\usepackage{multirow}
\usepackage[hidelinks]{hyperref} 

\begin{document}

\title{Wormholes exact solutions in high dimensions General Relativity}

\author{I. A. Sarmiento-Alvarado}
    \email{ignacio.sarmiento@cinvestav.mx}
\author{Leonel Bixano}
    \email{leonel.delacruz@cinvestav.mx}
\author{Tonatiuh Matos}%
 \email{tonatiuh.matos@cinvestav.mx}
\affiliation{Departamento de F\'{\i}sica, Centro de Investigaci\'on y de Estudios Avanzados del Intituto Politécnico Nacional, Av. Intituto Politécnico Nacional 2508, San Pedro Zacatenco, M\'exico 07360, CDMX.
}%

\date{\today}

\begin{abstract}
In the present work, we develop and examine a series of exact solutions to Einstein's 5-dimensional field equations in the vacuum, which depend on two constant parameters, $p$ and $q$, which generalize the solutions of Lü and Mei \cite{LU2008511} belonging to our class $p=2$. This category of solutions can be split into two sections: when $p$ is odd, it represents a compact object that may have naked singularities. However, the intriguing outcome occurs when $p$ is even, as asymptotically Ricci flat wormholes emerge in this scenario.
The Kreschman invariant of these solutions depends on the constant parameter $l = 1 + \frac{3 q^2 - p^2}{4}$.
When $l = 0$ and $l \leq -\frac{1}{2}$, the solutions are regular.
For the specific cases where $l \leq 0$, or $l > 0$ such that $q < 0$ for $p \geq 2$, $q \leq \frac{1}{3}$ for $p  = 2$, and $q \leq \frac{1 + \sqrt{2}}{3}$ for $ p \geq 4$, this class of wormholes adheres to Wormhole Cosmic Censorship, implying that the throat effectively obscures all causal anomalies and singularities.
In our analysis, we investigated the embedded geometry, geodesics, singularities, potential event horizons, ergoregions, and the wormhole throat.
\end{abstract}

\maketitle

\section{Introduction}  %

Wormholes have recently gained greater prominence and have been investigated within the framework of four-dimensional General Relativity, applicable to both general relativity and its extensions. In the five-dimensional realm, there is also interest in using geometry to clarify certain exotic phenomena, such as the EPR=ER conjecture. For instance, the Dzhunushaliev model \cite{Dzhunushaliev:1998rz} proposes a wormhole-like solution within five-dimensional Kaluza–Klein theory, where a \textit{naked charge} connects two four-dimensional Reissner–Nordström regions. In this scenario, electric flux enters one universe and exits through the other, while a foam of virtual wormholes adjusts the observed charge on a larger scale. In his subsequent research, the same author investigates “flux tube” type configurations and string-like solutions, supporting the concept that non-trivial topology in 5D can be perceived from a 4D perspective \cite{Dzhunushaliev:1998rz}.
In the study by Lü and Mei \cite{LU2008511}, they construct a series of Ricci-flat Lorentzian wormholes in dimensions $D \geq 5$. Specifically, in the case of $D = 5$, these wormholes can be viewed as tachyonic string states because their linear momentum surpasses their tension. Although they are impassable by time-like or null geodesics, they permit traversal with minimal acceleration and facilitate the quantum tunnelling of scalar fields. When these wormholes are incorporated into a 5D supergravity framework along with gauge fields, they can acquire a charge, linking an $AdS_3\times S^2$ terminus to a flat Minkowski terminus, with throat properties influenced by gauge parameters.

In this case, by applying the method proposed in \cite{Sarmiento-Alvarado2025}, we derive a family of solutions that extends the ones presented in \cite{Dzhunushaliev:1998rz,LU2008511}. Specifically, when we assign the parameters $p=2$ and $q=0$ to our solution, it aligns with the aforementioned solutions. Moreover, when $p$ is even, a sub-family emerges that can be considered as asymptotically Ricci-flat wormholes, which are further divided into regular wormholes (without singularities) and non-regular ones. Furthermore, for specific parameter values of $p$ and $q$, the wormhole's throat hides all the causal irregularities of the topology, adhering to the Wormhole Cosmic Censorship Conjecture introduced in \cite{Matos:2012gj} and explained in \cite{axioms14110831}.

The structure of the paper is as follows.
In Section \ref{section:gensol}, we build solutions to the EFE.
The general analysis of these solutions carried out in Section \ref{section: struct sols}.
In Section \ref{section: p2}, we study a particular solution.
Finally, in Section \ref{section: conclusions}, we present some conclusions.

\section{General solution}
\label{section:gensol}  %

In this section, we will build an exact solution to the 5-dimensional EFE using the technique described in appendix \ref{Apendice: Method Explanation} along with the mathematical tools from \cite{Sarmiento-Alvarado2023,Sarmiento-Alvarado2025}.

Let
\begin{equation}
    A_1 = \left[\begin{array}{rrr}
            1 &0&0 \\
           0 & 1 &0 \\
            0&0& -2 \\ 
    \end{array}\right]
    \text{ and }
        A_3 = \left[\begin{array}{rrr}
           0 & -1 &0 \\
            1 &0&0 \\
           0 &0& 0
        \end{array}\right]
\end{equation}
be a pair of pairwise commuting matrices of $\mathfrak{sl} ( 3, \mathbb{R} )$.
The exponential of $\xi^a A_a$ is
\begin{equation}
    e^{ \xi^a A_a } = \left[
        e^{ \xi^1 }
        \left[\begin{array}{rrrr}
            \cos \xi^2 & -\sin \xi^2 \\
            \sin \xi^2 & \cos \xi^2 \\
        \end{array}\right],
        e^{ -2 \xi^1 }
    \right].
\end{equation}
If $g_0$ is a constant matrix in $\mathcal{I} (\{ A_1, A_2 \})$, then it has the form
\begin{equation}
    g_0 = - \operatorname{diag} \left[
        \left[\begin{array}{rr}
            C_t & D \\
            D & -C_t \\
        \end{array}\right],
        C_\phi
    \right] ,
\end{equation}
where $C_t, D, C_\phi$ are real constants.
The generalized Laplace equation,
\begin{equation}
\label{Laplace eq Boyer-Lindquist coordinates}
    ( ( r^2 + r_0^2 ) \xi^a_{, r} ) _{, r} + \frac{1}{\sin \theta} ( \xi^a_{, \theta} \sin \theta ) _{, \theta} = 0 ,
\end{equation}
in terms of the Boyer-Lindquist coordinates,
\begin{equation}
\label{def Boyer-Lindquist coordinates}
    \alpha = \sqrt{ r^2 + r_0^2 } \sin \theta
    \text{ and }
    \zeta = r \cos \theta ,
\end{equation}
has the solution 
\begin{equation}
\label{gen Laplace eq sol}
    \xi^a ( r, \theta )
    = \mathscr{C}^a \ln \alpha 
    + \frac{\mathscr{D}^a}{r_0} \arctan \left( \frac{r}{r_0} \right)
    + \xi^a_0 ,
\end{equation}
where $r \geq 0$,
$0 \leq \theta \leq \pi$,
$r_0$ is a positive constant and
$\mathscr{C}^a, \mathscr{D}^a, \xi^a_0$ are real constants.
To obtain a solution to the chiral equation, we set $\mathscr{C}^1 = -\frac{2}{3}$, $\mathscr{C}^2 = 0$, $\mathscr{D}^1 = r_0 q$ and $\mathscr{D}^2 = r_0 p$, where $p$ and $q$ are real constants.
Therefore,
\begin{equation}
    g
    = -\rho^{-\frac{2}{3}} \operatorname{diag} \left[
        \Xi \left[\begin{array}{rr}
             U & V \\
             V & -U
        \end{array}\right] ,
        \frac{ \rho^2 \Xi^{-2} }{ \mu^2 + \nu^2 }
    \right] ,
\end{equation}
where
\begin{subequations}
\begin{align}
\Phi_r (r)
&   = \arctan \left( \frac{r}{r_0} \right) ,
\\    \Xi (r)
&   = \exp\left( q \left( \Phi_r - \frac{\pi}{2} \right)\right) ,
\\  U (r)
&   = \mu \cos( p \, \Phi_r) - \nu \sin( p \, \Phi_r),  
\\  V (r)
&   = \nu \cos( p \, \Phi_r) + \mu \sin (p \, \Phi_r) ,
\end{align}
\end{subequations}
$\mu = C_t \cos \xi^2_0 - D_t \sin \xi^2_0$, $\nu = D_t \cos \xi^2_0 + C_t \sin \xi^2_0$.

In Boyer-Lindquist coordinates, Eq. \eqref{SL invariant field eq f} changes to
\begin{equation}\label{diff eq f Boyer-Lindquist coordinates}
\begin{aligned}
    \left( \ln f \alpha^{1-1/n} \right)_{, r} & = \frac{\operatorname{tr} A_a A_b}{4} \frac{
        r \sin ^2 \theta
    }{ \Delta } \Bigg[
        \xi^a_{, r} \xi^b_{, r} 
\\&     + \frac{
            2 \cot \theta \xi^a_{, r} \xi^b_{, \theta}
        }{r}
        - \frac{
            \xi^a_{, \theta} \xi^b _{, \theta}
        }{ r^2 + r_0^2 }
    \Bigg] ,
\\  \left( \ln f \alpha^{1-1/n} \right)_{, \theta}
&   = - \frac{\operatorname{tr} A_a A_b}{4} \frac{
        \sin \theta \cos \theta
    }{ \Delta } \Bigg[
        \xi^a_{, r} \xi^b_{, r}
   \\& - \frac{
            2 r \tan \theta \xi^a_{, r} \xi^b_{, \theta}
        }{
            r^2 + r_0^2
        }
    - \frac{
            \xi^a_{, \theta} \xi^b_{, \theta}
        }{
            r^2 + r_0^2
        }
    \Bigg] ,
\end{aligned}
\end{equation}
where
\begin{equation}
    \Delta = \frac{
        r^2 + r_0^2 \cos^2 \theta
    }{ r^2 + r_0^2 } \equiv \frac{\Delta_\theta}{\Delta_r} . 
\end{equation}

For the solutions \eqref{gen Laplace eq sol}, $f$ is given by
\begin{equation}
    f  ( r, \theta )
    = \Delta^\frac{3 q^2 - p^2}{4}
    \Xi^{-2} .
\end{equation}

Therefore, a solution to the EFE is
\begin{equation}\label{general sol}
\begin{aligned}
    \hat{g}
&   = \frac{
        \Delta^l (
        dr^2
        + \Delta_r d\theta^2
    )
    + \Delta_r \sin^2 \theta d\phi^2
    }{\Xi^2} 
\\& + U \Xi ( d\psi^2 - dt^2 )
    + 2 V \Xi dt d\psi ,
\end{aligned}
\end{equation}
where $l = 1 + \frac{3 q^2 - p^2}{4}$.

An interesting case is when $p$ is a natural number.
In this case, the functions $\cos( p \,\Phi_r)$ and $\sin (p \, \Phi_r)$ are related to Chebyshev polynomials as $\cos (p \,\Phi_r) = T_p ( \cos \,\Phi_r )$ and $\sin (p \,\Phi_r) = \sin \,\Phi_r U_{p - 1} ( \cos \,\Phi_r )$, where $T_p$ and $U_{p - 1}$ are Chebyshev polynomials of the first and second kind, respectively.

Then, taking $p\in \mathbb{N}$ into account, we formulate our solution, which we shall examine comprehensively:
\begin{equation}\label{Solucion p q generales}
\begin{aligned}
    \hat{g}_p
&   = \frac{
        \Delta^l (
        dr^2
        + \Delta_r d\theta^2
    )
    + \Delta_r \sin^2 \theta d\phi^2
    }{\Xi^2} 
\\& + \mathscr{U}_p \Xi \Big( 
        d\psi^2 -dt^2 \Big)
    + 2 \mathscr{V}_p \Xi dt d\psi,
\end{aligned}
\end{equation}
where
\begin{align}
    \mathscr{U}_p ( r )
&   = \mu T_p \left( \frac{r_0}{\sqrt{ r^2 + r_0^2 }} \right)
    - \frac{\nu r}{\sqrt{ r^2 + r_0^2 }} U_{p - 1} \left( \frac{r_0}{\sqrt{ r^2 + r_0^2 }} \right),
\\  \mathscr{V}_p ( r )
&   = \nu T_p \left( \frac{r_0}{\sqrt{ r^2 + r_0^2 }} \right)
    + \frac{\mu r}{\sqrt{ r^2 + r_0^2 }} U_{p - 1} \left( \frac{r_0}{\sqrt{ r^2 + r_0^2 }} \right) .
\end{align}

\subsection{Asymptotic behaviour of the metric} %

We can obtain the asymptotic behavior ($r \rightarrow +\infty$) of the metric functions corresponding to \eqref{Solucion p q generales}, which are:

\begin{subequations}\label{Asimptotic total}
    \begin{align}
        &g_{tt}\rightarrow\nu  \sin \left(\frac{\pi  p}{2}\right)-\mu  \cos \left(\frac{\pi  p}{2}\right),\\
        &g_{rr} \rightarrow 1,\\
        &g_{\theta \theta} \rightarrow r^2,\\
        &g_{\phi \phi} \rightarrow r^2 \sin^2{\theta},\\
        &g_{\psi \psi}\rightarrow \mu  \cos \left(\frac{\pi  p}{2}\right)-\nu  \sin \left(\frac{\pi  p}{2}\right), \\
        &g_{t \psi} \rightarrow \mu  \sin \left(\frac{\pi  p}{2}\right)+\nu  \cos \left(\frac{\pi  p}{2}\right) \label{Asimptotic g5t}.
    \end{align}
\end{subequations}

The criterion for asymptotic flatness within the context of five-dimensional spaces, which will be employed in our analysis, is
\begin{equation}
\label{5D Minkoswki metric}
    \hat{g}_p \to -dt^2+dr^2+r^2(d\theta^2+\sin^2{\theta}d\phi^2)+d\psi^2
\end{equation}
as $r \to +\infty$.
In Eqs. \eqref{Asimptotic total}, two significant terms appear: the function $\sin$ and $\cos$, along with two notable facts:
\begin{center}
\begin{tabular}{l}
    if $p$ is \textbf{even}, then $\sin \left({\frac{p\pi}{2}} \right) = 0$ and $\cos \left({\frac{p\pi}{2}}\right) =(-1)^\frac{p}{2}$ ,
\\  if $p$ is \textbf{odd}, then $\cos \left({\frac{p\pi}{2}} \right)=0$ and $\sin \left({\frac{p\pi}{2}}\right) =(-1)^{\frac{p-1}{2}}$ .
\end{tabular}
\end{center}

To achieve an asymptotically flat metric, it is necessary that $g_{tt} \to -1$, $g_{\psi \psi} \to +1$ and $g_{t \psi} \to 0$ as $r \to +\infty$.
Therefore, $\mu=(-1)^\frac{p}{2}$ and $\nu=0$ if $p$ is \textbf{even}, while if $p$ is \textbf{odd}, then $\mu=0$ and $\nu=(-1)^{\frac{p+1}{2}}$.

\subsection{Extension to negative values of $r$}    %

The solution presented in \eqref{Solucion p q generales} is not only applicable to $\{r_0,r\}>0$; it is also relevant for values in $\{r,r_0\}\in \mathbb{R}$.
Specifically, the solution, initially developed for compact objects, can be extended to encompass wormholes due to its consistency across both negative and positive values of the variables involved. However, for practical application, it is essential to consider the \textit{assymptotic behaviour} of the metric to ensure consistency with $\hat{g}_p \to -dt^2+dr^2+r^2(d\theta^2+\sin^2{\theta}d\phi^2)+d\psi^2$ as $r \to -\infty$.

The asymptotic behavior of \eqref{Solucion p q generales} with respect to $r \to -\infty$ has been determined:
\begin{subequations}\label{Asimptotic total Negativo}
    \begin{align}
        &g_{tt} \rightarrow -e^{-\pi  q} \left(\mu  \cos \left(\frac{\pi  p}{2}\right)+\nu  \sin \left(\frac{\pi  p}{2}\right)\right), \label{gtt asymptotic negativo}\\
        &g_{\psi \psi} \rightarrow e^{-\pi  q} \left(\mu  \cos \left(\frac{\pi  p}{2}\right)+\nu  \sin \left(\frac{\pi  p}{2}\right)\right), \label{gpspsi asymptotic negativo}\\
        &g_{rr} \to e^{2 \pi  q},\\
        &g_{\theta \theta} \to e^{2 \pi  q} r^2,\\
        &g_{\phi \phi} \rightarrow e^{2 \pi  q} r^2 \sin^2 \theta,\\
        &g_{t \psi} \to 0.
    \end{align}
\end{subequations}
Through the rescaling transformations $r \to e^{\pi q} r$, $\psi \to e^{-q \frac{\pi}{2}} \psi$ and $t \to e^{-q \frac{\pi}{2}} t$, we obtain the metric \eqref{5D Minkoswki metric} if $p$ is even.
However, for odd $p$, we have $\hat{g}_p = dt^2+dr^2+r^2(d\theta^2+\sin^2{\theta}d\phi^2)-d\psi^2$, this scenario would result in a violation of causality if the variable $\psi \sim \psi +2n \pi$.

\textit{Therefore, the sole feasible solutions pertaining to wormholes are those which possess $p$ as an even parameter.}

\section{Structure of the solutions}    %
\label{section: struct sols}

\subsection{Singularities}  %

Prior to initiation, it is imperative to acknowledge that in $5$-D, all components of the Ricci tensor are zero, i.e., $R_{A B} = 0$, followed by $R = 0$. Consequently, analogous reasoning in a five-dimensional context indicates that this solution inherently satisfies the Null Energy Condition \textit{NEC}.

Taking into account the \textit{general solution} \eqref{general sol}, it is possible to compute the \textbf{Kreschman invariant}, denoted by $K$, for arbitrary $p$ and $q$.
For $l = 0$, we have
\begin{equation}
\label{Kreschman invariant l zero}
        K = \frac{
            r_0^2 \mathcal{F}_1 \Xi^4
        }{ 
            8 \Delta_r^4
        }  ,    
\end{equation}
where $ \mathcal{F}_1 = - \frac{64}{3} p^{4} r_{0}^{2}
+ 128 p^{2} q r_{0} r
+ \frac{128}{3} p^{2} r_{0}^{2}
- 384 r^2
+ 640 q r_{0} r
+ \frac{1088}{3} r_{0}^{2} $.
Since $\Delta_r > 0$ on $\mathbb{R}$, then $K$ does not have any singularities.
\textit{Therefore, the solutions with $l = 0$ are regular}.
The metrics $\hat{g}_p$ with $\vert q \vert = \sqrt{\frac{ p^2 - 4 }{3}}$ for all $p \geq 2$ are regular.

If $l \neq 0$, $K$ is given by
\begin{equation}
\label{Kreschman invariant}
        K = \frac{ r_0^2
            \mathcal{F}_0 \, \, \Delta_r^{\,\,2 l - 4} \, \, \Xi^4
        }{ 
            8 \, \Delta_\theta^{\,\,2 l + 1}
        }  ,    
\end{equation}
in which $\mathcal{F}_0$ (For a comprehensive expression, refer to \eqref{eq:F0}) represents a polynomial function of $r$.
It is essential to note that $\Delta_\theta = 0$ at $r = 0$, $ \theta = \frac{\pi}{2}$.
Thus, $K$ exhibits singularities at $r = 0$, $\theta = \frac{\pi}{2}$, when $2l+1>0$ is considered. 
\textit{We conclude that for values $l \leq -\frac{1}{2}$ the solutions are regular}.
Consequently, the metrics $\hat{g}_p$ with $\vert q \vert \leq \sqrt{ \frac{p^2}{3} - 2 }$ for all $p \geq 3$ are regular.

\subsection{Event horizons} %
In order to determine the event horizons within these solutions, we shall employ the Killing vector $K_t=\partial_t + \Omega\partial_\psi$.
It is important to note that the parameter $\Omega$ must remain constant at the event horizon. Let
\begin{align*}
    &g(K_t,K_t)=g_{tt}+2\Omega g_{t\psi}+\Omega^2g_{\psi \psi}=0, \\
    & \Rightarrow \Omega_{\pm}=-\frac{g_{t\psi}}{g_{\psi \psi}}\pm \frac{\sqrt{ (g_{t\psi})^2 -g_{\psi \psi} g_{tt} }}{g_{\psi \psi}},
\end{align*}
and considering the non degenerate solution
\begin{subequations}
\begin{align}
    &\Omega=\Omega_-=\Omega_+=-\frac{g_{t\psi}}{g_{\psi \psi}},\\
    &g_{\psi \psi}g_{tt}-(g_{t\psi})^2 =0\label{Horizonte condicion},
\end{align}
\end{subequations}
we are able to determine the event horizon.
Using the metric \eqref{Solucion p q generales} into equation \eqref{Horizonte condicion}, it becomes evident that:
\begin{equation*}    
    (g_{t\psi})^2 - g_{\psi \psi}g_{tt}
   = \Xi \Big( \mathscr{U}_p^2+\mathscr{V}_p^2  \Big)
   = \Xi (\mu^2+\nu^2)
\end{equation*}
on $\mathbb{R}$.
If $\mu = \nu = 0$, then $g_{tt} = g_{\psi \psi} = g_{t \psi} = 0$ on $\mathbb{R}$.
Hence, at least one of them must be non-zero.
This implies that \textit{there are no event horizons for arbitrary $p$ and $q$}.
For the metrics $\hat{g}_p$ that describe wormholes, we have $\mu^2= 1$.

\subsection{Ergo-region}    %

In this section, and in the remainder of this article, we only consider the case where $p$ is an even number.

The other important region is the called ergoregion, and we will obtain taking into account 
\begin{equation}
    g ( \partial_t, \partial_t )
    = g_{tt}
    =0 ,
\end{equation}
which implies that $\mathscr{U}_p = 0$.
Thus, the roots $r_k$ of $\mathscr{U}_p$ satisfy $\Phi_{r_k} = \frac{2 k + 1}{p} \frac{\pi}{2}$, where $k \in \mathbb{Z}$.
Since the image of $\Phi_r$ on $\mathbb{R}$ is $\Phi_r ( \mathbb{R} ) = ( -\frac{\pi}{2} , \frac{\pi}{2} )$, then $\vert 2 k + 1 \vert < p$, so that  the roots of $\mathscr{U}_p$ are given by $r_k = r_0 \tan \frac{2 k + 1}{p} \frac{\pi}{2}$ for all $k \in \{ -\frac{p}{2}, \ldots, \frac{p}{2} - 1 \} \subset \mathbb{Z}$.
There are $\frac{p}{2}$ positive roots and $\frac{p}{2}$ negative roots.
Taking into account that $\Phi_r$ is an increasing function, the largest root $r_e = r_0 \cot \frac{\pi}{2 p}$ corresponds to $k = \frac{p}{2} - 1$.

Let $\Phi_r^\prime = p ( \frac{\pi}{2} - \Phi_r )$.
Then, the image of $\Phi_r^\prime$ on $I = ( r_e, \infty )$ is $\Phi_r^\prime (I) = ( 0, \frac{\pi}{2} )$.
Expressing $\mathscr{U}_p$ as $\mathscr{U}_p = \mu^2 \cos \Phi_r^\prime$, we obtain $\mathscr{U}_p > 0$ on $I$ as a consequence of the fact that $\cos \Phi_r^\prime > 0$ on $\Phi_r^\prime (I)$.
From \eqref{Solucion p q generales} we can see that $g_{tt}=-g_{\psi \psi}$, thus:
\begin{equation}\label{CondicionDeCausalidad}
    g(\partial_t,\partial_t)<0 \text{ and } g(\partial_\psi,\partial_\psi)>0 \text{ on } I.
\end{equation}

\textit{A crucial point to highlight is that, within the ergo-region where $r<r_e$, Closed Timelike Curves (CTCs) arise when considering $\psi \sim \psi +2n \pi$, thereby disrupting causality. Therefore, it is essential to uphold the condition \eqref{CondicionDeCausalidad} to prevent the existence of CTCs.}

\subsection{Geodesics}  %

We obtain the equations of motion from the Lagrangian 
\begin{equation}\label{Lagrangiano}
\begin{aligned}
    \mathscr{L}
&   = \frac{
        \Delta^l (
        \dot{r}^2
        + \Delta_r \dot{\theta}^2
    )
    + \Delta_r \sin^2 \theta \dot{\phi}^2
    }{2 \Xi^2} 
\\& + \Xi \mathscr{U}_p \frac{
        \dot{\psi}^2
        - \dot{t}^2
    }{2}
    + \Xi \mathscr{V}_p \dot{t} \dot{\psi} ,
\end{aligned}    
\end{equation}
where $\dot f \equiv d f /d\lambda$ and $\lambda$ is the affine parameter.
We consider solutions to equations of motion where both $\theta = \theta_0$ and $\phi=\phi_0$ are constant.
From the equations of motion for $t$ and $\psi$, we obtain
\begin{equation}
\begin{array}{l}
\label{velocities}
    \dot t = \frac{
        E \mathscr{U}_p
        + J \mathscr{V}_p
    }{\Xi} ,
\\  \dot \psi
    = \frac{
        J \mathscr{U}_p
        - E \mathscr{V}_p
    }{\Xi} ,
\end{array}
\end{equation}
where $p_t = \mathscr{L}_{, \dot{t}} = -E$, $p_\psi = \mathscr{L}_{, \dot{\psi}} = J$, $\mathscr{L}_{, \dot{\phi}} = p_\phi$ represent the integration constants, since the Lagrangian \eqref{Lagrangiano} possesses three cyclic variables ($t,\psi,\phi$).
In this context, $E$ denotes the energy, while $(p_\phi,J)$ is the angular momentum of the particle along the $(\phi,\psi)$-direction, respectively.
The equation of motion for $r$ with $\theta,\phi$: constant, is
\begin{equation}
\label{geodesic eq}
    \frac{\Delta^l}{\Xi} \dot{r}^2
    = ( E^2 - J^2 ) \mathscr{U}_p
    + 2 E J \mathscr{V}_p .
\end{equation}

Now, we write $\mathscr{V}_p$ as a function of $\Phi_r^\prime$, $\mathscr{V}_p = - \mu^2 \sin \Phi_r^\prime$.
Since $\sin \Phi_r^\prime > 0$ on $\Phi_r^\prime (I)$, we have $\mathscr{V}_p < 0$ on $I$.
$\mathscr{V}_p$ takes the value -1 at $r = r_e$.

We assume that $E > 0$.
Consequently, $\dot t > J \mathscr{V}_p \Xi^{-1}$ on $I$, where we have taken the fact that $\mathscr{U}_p > 0$ on $I$.
The criterion $\dot{t} > 0$ precludes velocities exceeding that of light.
It holds if $J \leq 0$, because $\mathscr{V}_p < 0$ on $I$.
Using it, we get $\frac{\Delta^l}{\Xi} \dot{r}^2 \geq ( E^2 - J^2 ) \mathscr{U}_p$.
The inequality $\dot r^2 > 0$ is satisfied on $I$ when $E^2 > J^2$.
Therefore, $E \geq -J \geq 0$.
On the other hand, we have the inequalities $\dot{t} \geq 0$ and $\dot{r}^2 \geq 0$ at $r = r_e$.
Equality holds when $J = 0$.

\subsection{Wormholes throat}   %
\label{throat}

In this section, we will determine the wormhole throat for the metrics $\hat{g}_p$ with even $p$.
Hochberg and Visser provide a definition of a generic static throat \cite{Hochberg:1997wp, Visser:1997yn}, while Simmonds and Visser give a definition in five dimensions in \cite{Simmonds:2025tet}.

In order to obtain the wormhole throat, we minimize the volume
\begin{equation}
    V (r) = 4 \pi^2 \int_0^\pi \sqrt{\gamma} d \theta
\end{equation}
of the 3-dimensional hypersurface $\Sigma$ given by $r = constant$, where $\gamma = \Xi^{-3} \Delta^l \mathscr{U}_p \Delta_r^2 \sin^2 \theta$ is the determinant of the induced metric on $\Sigma$.
The unit vector $ n^r = \frac{\Xi}{\sqrt{\Delta^l}} $ is normal to $\Sigma$ and its dual is $ n_r = \frac{\sqrt{\Delta^l}}{\Xi} $.

Let $m$ be a positive constant.
We define $F ( x, m ) = \cot \frac{\pi}{2 x} - m x$ for all $x > 1$.
To determine $m$, we impose $F ( 4, m ) \geq 0$.
Hence, $m \leq \frac{1 + \sqrt{2}}{4}$.
Differentiating $F$ and using the inequality $\sin \frac{\pi}{2 x} < \frac{\pi}{2 x}$ for all $x > 1$, we obtain $F_{, x} > 0$.
This means that $F$ is a strictly increasing function for all $x > 1$.
Therefore, $F \geq 0$ for all $x \geq 4$.

The derivative of $V$ is given by
\begin{equation}
    V_{, r}
    = 4 \pi^2 \int_0^\pi \sqrt{\gamma} \mathscr{K} n_r d \theta ,
\end{equation}
where
\begin{align}
    \mathscr{K}
&   = \frac{ \partial }{\partial n} ( \ln \sqrt{\gamma} )
\\& = \frac{n^r r_0}{2 \Delta_r} \left(
        4 \frac{r}{r_0}
        + 2 l \frac{r}{r_0} \left( \frac{1}{\Delta} - 1 \right)
        - p \frac{ \mathscr{V}_p }{ \mathscr{U}_p }
        -3 q
    \right)
\end{align}
is the mean curvature \cite{Gourgoulhon2012} and $\frac{\partial}{\partial n}$ is the normal derivative.
If $\mathscr{K} > 0$ on $I$, then $V_{, r} > 0$, so that $V$ is a strictly increasing function.
Given that $0 = V (r_e) < V (r)$ on $I$, $V (r_e)$ is a minimum.
This means that the throat is at $R_G = r_e$.

When $l = 0$, $q$ is given by $\vert q \vert = \sqrt{\frac{p^2 - 4}{3}}$ for all $p \geq 2$.
Since $\mathscr{U}_p > 0$ and $\mathscr{V}_p < 0$ on $I$, $\mathscr{K}$ verifies the inequality
\begin{equation}
\label{ineq mean curvature non neg}
    \mathscr{K} > \frac{2 \Xi r_0}{\Delta_r} \left(
        \cot \frac{\pi}{2 p}
        - \frac{3}{4} q
    \right) .
\end{equation}
Using $q < \frac{p}{\sqrt{3}}$ for all $p \geq 4$, we get $\mathscr{K} > \frac{2 \Xi r_0}{\Delta_r} F ( p, \frac{\sqrt{3}}{4} ) > 0$.
If $p = 2$, then $q = 0$, so that $\mathscr{K} > 0$.

Given that $0 \leq \cos^2 \theta \leq 1$, it follows that $r^2 \leq \Delta_\theta \leq \Delta_r$, and therefore $0 \leq \frac{1}{\Delta} - 1 \leq \frac{r_0^2}{r^2}$ for all non-zero $r$.
If $l < 0$, then $\vert q \vert < \sqrt{\frac{p^2 - 4}{3}}$ for all $p \geq 4$, so that
\begin{equation}
    \mathscr{K} > \frac{2 \Xi r_0}{\Delta_r} \left(
        F ( p, \tfrac{\sqrt{3}}{4} ) \cot \frac{\pi}{2 p}
        + \frac{l}{2}
    \right) \tan \frac{\pi}{2 p}.
\end{equation}
After some algebraic manipulation, we can write the sum on the right-hand side of the above inequality as $\left( F ( p, \tfrac{\sqrt{3}}{8}) + \frac{\sqrt{ 3 p^2 - 32 l}}{8} \right) \left( F ( p, \tfrac{\sqrt{3}}{8}) - \frac{\sqrt{ 3 p^2 - 32 l}}{8} \right)$.
Inequality $\sqrt{ 3 p^2 - 32 l} < \sqrt{11} p$ implies that $F ( p, \tfrac{\sqrt{3}}{8}) - \frac{\sqrt{ 3 p^2 - 32 l}}{8} > F ( p, \tfrac{ \sqrt{3} + \sqrt{11} }{8} ) > 0$.
Therefore, $\mathscr{K} > 0$ for all $p \geq 4$. 
Note that $l$ is non-negative for $p = 2$.

$\mathscr{K}$ fullfils Ineq. \eqref{ineq mean curvature non neg} for $l > 0$.
Then, $q$ verifies $\sqrt{\frac{p^2 - 4}{3}} < \vert q \vert$ for all $p \geq 2$.
When $q < 0$, we have $\mathscr{K} > 0$ for all $p \geq 2$.
For the case $q > 0$, we impose $q \leq \frac{1 + \sqrt{2}}{3} p$ for all $p \geq 4$ and $q \leq \frac{1}{3}$ for $p = 2$, which implies $\mathscr{K} > 0$.
We need to use the following method to find the minimum of a real function of one real variable for the values of $q$ that do not satisfy these upper bounds:
\begin{enumerate*}
\item   Find all critical points $r_c$ of $V$, that is, $V_{, r} (r_c) = 0$.
\item   Determine whether $V_{, r r} (r_c) > 0$ holds.
\end{enumerate*}
$V_{, r r}$ is given by
\begin{equation}
    V_{, r r}
    = 4 \pi^2 \int_0^\pi \sqrt{\gamma} n_r^2 \left(
        \mathscr{K}^2
        + \frac{\partial \mathscr{K}}{\partial n}
        + a \mathscr{K} 
    \right) d \theta ,   
\end{equation}
where $a = \frac{\partial}{\partial n}( \ln \sqrt{g_{rr}} )$.

\section{\texorpdfstring{Solution $p=2$}{Values p=2}}   %
\label{section: p2}

In this section, we study the metric $\hat{g}_2$.
For this case, $\mu = -1$ and $\nu = 0$.

\begin{figure}[b]
        \includegraphics[width=0.48\textwidth]{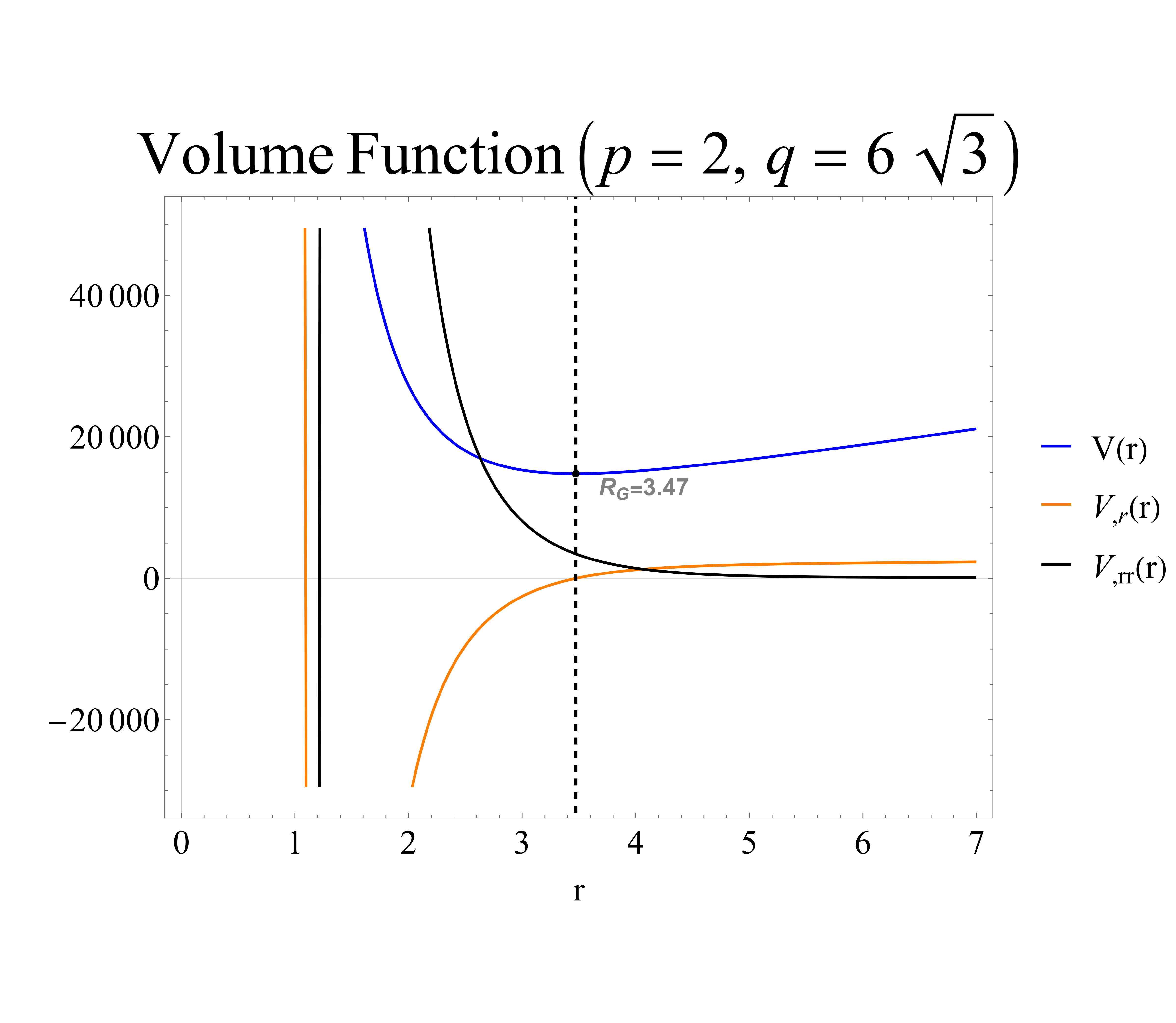}
        \caption{The volume function, along with its first and second derivatives, has been determined through numerical methods for the parameters $p=2$, $q=6\sqrt{3}$, and $r_0=1$. By applying these numerical methods, it is found that in this scenario, the throat is $R_G=3.47197$ ($V_{,r}(R_G)=0$ and $V_{,rr}(R_G)>0$) that corresponds to dashed black line.}
        \label{fig:RFTotal}
\end{figure}

In \cite{LU2008511,Dzhunushaliev:1998rz}, the authors study the metric $\hat{g}_2$ with $q = 0$.
Dzhunushaliev found that $R_G = r_0$ \cite{Dzhunushaliev:1998rz}.
In Subsection \ref{throat}, we find $R_G = r_e$ for $q \leq \frac{1}{3}$.
However, we cannot prove that $R_G = r_e$ holds when $q > \frac{1}{3}$.
We determine numerically $R_G=3.47197$ for $q=6\sqrt{3}$.
In Figure \ref{fig:RFTotal}, we plot the behaviors of $V$, $V_{, r}$ and $V_{, r r}$ with respect to $r$.

\subsection{Embedded diagrams}

To analyse the geometry of these objects, we will construct an embedded diagram employing a hyper-surface on which either $\{x^5,t,\theta\}$ or $\{x^5,t,r\}$ is held constant in (\ref{Solucion p q generales}), following the methodology presented in the articles by \cite{Bixano:2025jwm,Bixano:2025bio,DelAguila:2015isj}. This approach will provide us with the profile ($r,\phi$) of the compact object or the shape ($\theta,\phi$).

The following differential equations yield the associated embedding diagrams in 5 Dimensions
\begin{subequations}\label{Ec de la GoemtriaHipersup general}
    \begin{align}
        &\overline{\rho}(u,v_0)^2= g_{hh}  (u,v_0), \label{rho(x) vcote} \\
        &\left( \frac{d\, \, \overline{\rho}}{du} \right)^2 +\left( \frac{dz}{du} \right)^2= g_{uu}(u,v_0). \label{EcDif rhoZ(x) vcte}
    \end{align}
\end{subequations}
where $h=\phi,\psi$ these two variables correspond to periodic variables, $u=r,\theta$ refers to the variable with respect to which we are embedding, and $v_0=\theta_0,r_0$ represents the corresponding constant dual variable, whose behaviour we aim to analyse with respect to $u$.

Taking into account the solution (\ref{Solucion p2 q0}), we can numerically solve equations (\ref{Ec de la GoemtriaHipersup general}), fixing the initial condition $z(0)=0$ for both $(u=r,v_0=\theta_0)$ and $(u=\theta,v_0=r)$.

\begin{figure}[b]
    \centering
    \begin{minipage}{0.48\textwidth}
    \centering
        \includegraphics[width=\textwidth]{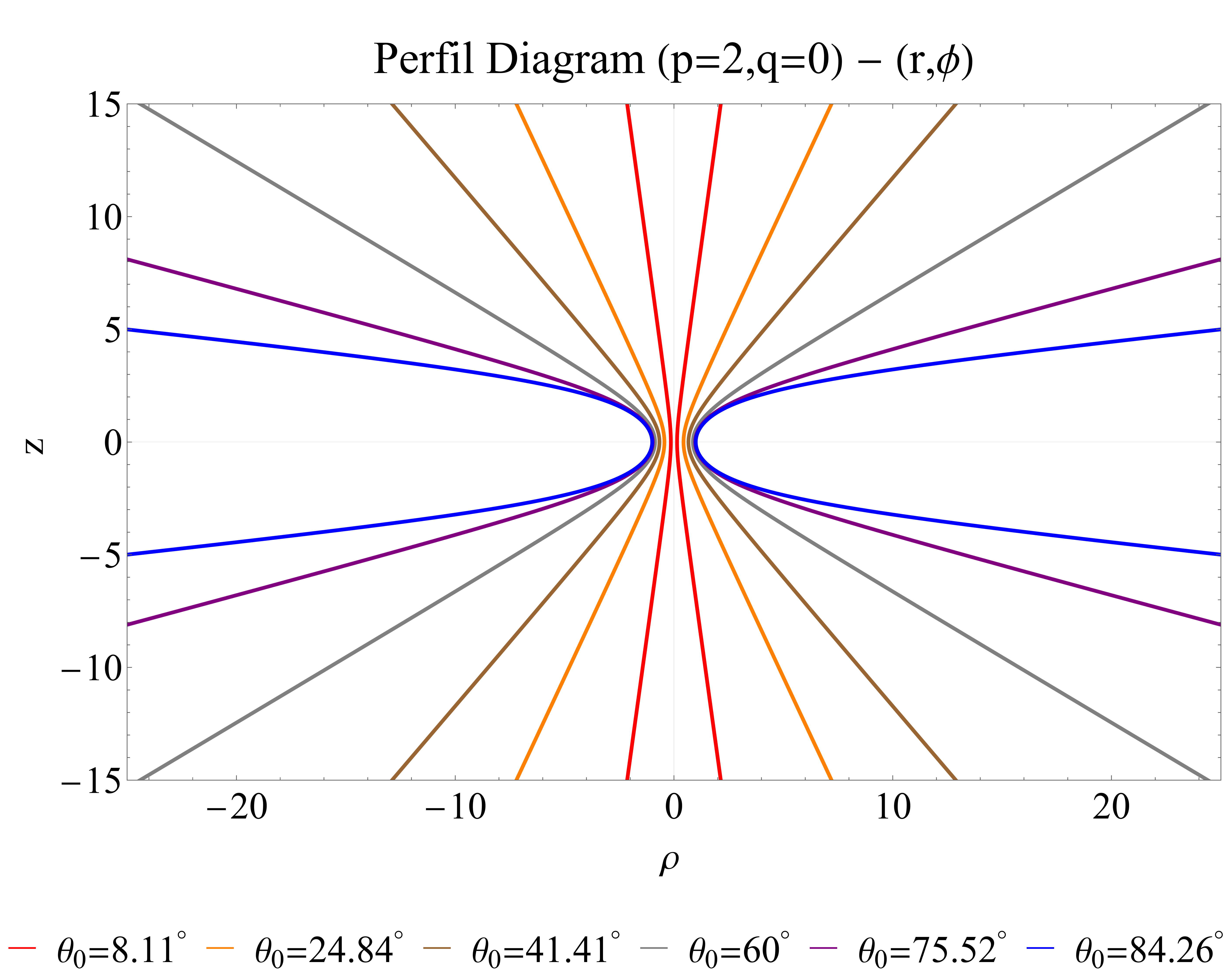}
        \subcaption{The embedding associated with $h=\phi,u=r,v_0=\theta_0$ pertains to the profile diagram of the compact object. Reference $z>0$ corresponds to one universe, while reference $z<0$ may refer either to another universe or potentially the same one. Each colour line represents a distinct constant angle.}
        \label{fig:CombinacionrPhip2q0}
    \end{minipage}
    \hfill
    \begin{minipage}{0.48\textwidth}
    \centering
    \includegraphics[width=\textwidth]{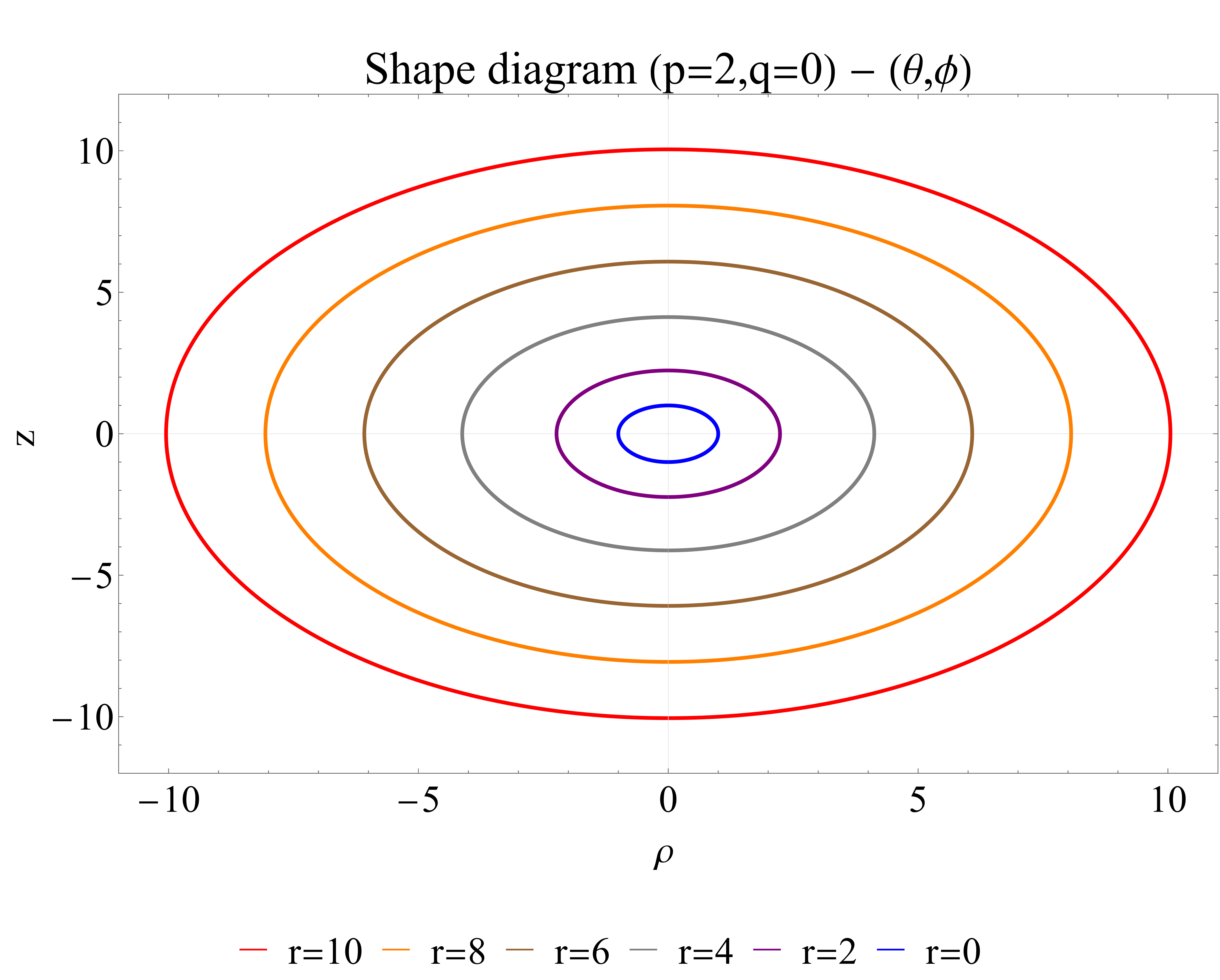}
        \subcaption{The embedding associated with $h=\phi,u=\theta,v_0=r_0$ pertains to the shape diagram of the compact object. Each colour line represents a distinct constant radius.}
        \label{fig:CombinacionThetaPhip2q0}
    \end{minipage}
    \caption{The embedding diagram corresponding to the solution referenced in \eqref{Solucion p2 q0}, the parameter was $r_0=1$.}
\end{figure}

\subsubsection{\texorpdfstring{Plotting for $q=0$}{Values q=0}}

To depict null geodesics, we will employ the Hamiltonian derived from the solution referenced in \eqref{Solucion p2 q0}:

\begin{equation}    
    H
    =\frac{E^2-J^2}{2} \mathscr{U}_p
    + E J \mathscr{V}_p
    + \frac{\csc ^2(\theta ) p_{\phi }^2}{2 \Delta_r}
    + \frac{p_{\theta }^2}{2 \Delta_r}
    + \frac{p_r^2}{2},
\end{equation}

\begin{figure}[b]
        \includegraphics[width=0.48\textwidth]{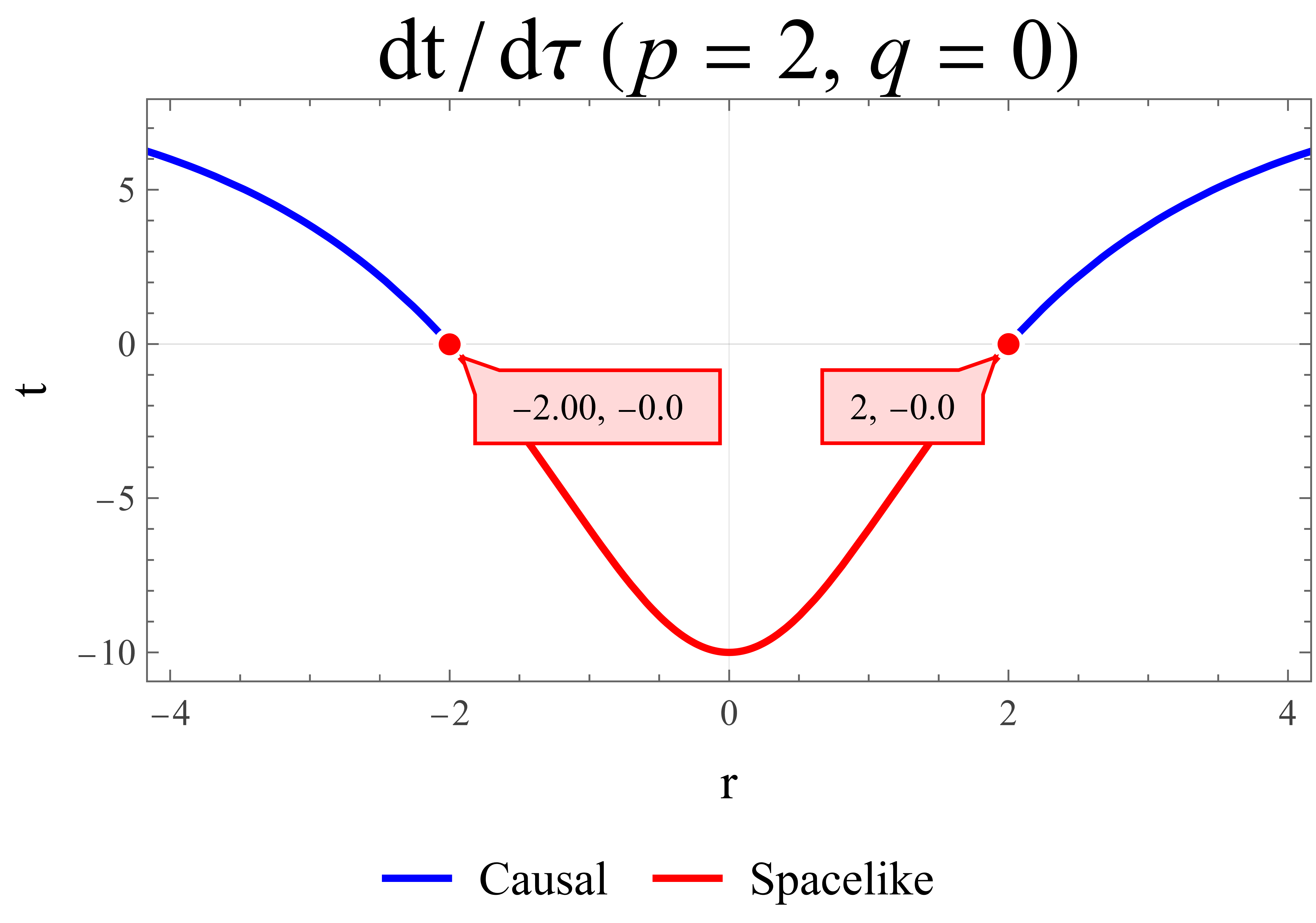}
        \subcaption{The behavior of $\dot{t}$, fixing the constants $J=0$, $E=10$, and $r_0=2$, reveals $\dot{t}>0$ for all values of $r>r_0$. Consequently, causality violations occur when $\dot{\psi}=-\dot{t}>0$. In other words, the red line represents a violation of causality and spacelike geodesics, whereas the blue region indicates causal geodesics. }
        \label{fig:Graficatpunto}
\end{figure}

Similarly, it is necessary to fulfil the condition $\dot{t} >0$ to ensure that causality is not violated and to prevent exceeding the speed of light, refer to \textbf{Figure }\ref{fig:Graficatpunto} for comprehensive visual elucidations.

To accurately plot null geodesics, it is essential to confirm that condition $H (x^\mu , p^\mu)=0$ is satisfied. Subsequently, employing numerical methods, we derive the following initial conditions fixing $r_0=R_G=2$, $E=10$, $J=0$, $p_\phi =1$:

\begin{align}
     r(0)=10, \, \theta(0)=\pi/12, \, p_\theta(0)=1 \, \Rightarrow \, p_r(0)\approx -9.59972, \label{Condiciones iniciales p=2 pi12} \\
     r(0)=10, \, \theta(0)=\pi/2, \, p_\theta(0)=1 \, \Rightarrow \, p_r(0)\approx -9.60669 \label{Condiciones iniciales p=2 pi2}.
\end{align}

The equations governing motion are systematically derived from Hamilton's equations, which are expressed by

    \begin{equation} \label{Ecuaciones de Hamilton}
        \dot{x}^{\mu} \equiv \frac{\partial \mathcal{H}}{\partial p_{\mu}},\qquad 
        \dot{p}_{\mu} \equiv -\frac{\partial \mathcal{H}}{\partial x^{\mu}}.
    \end{equation}

\begin{figure}[b]
    \centering
    \begin{minipage}{0.48\textwidth}
    \centering
    \includegraphics[width=\textwidth]{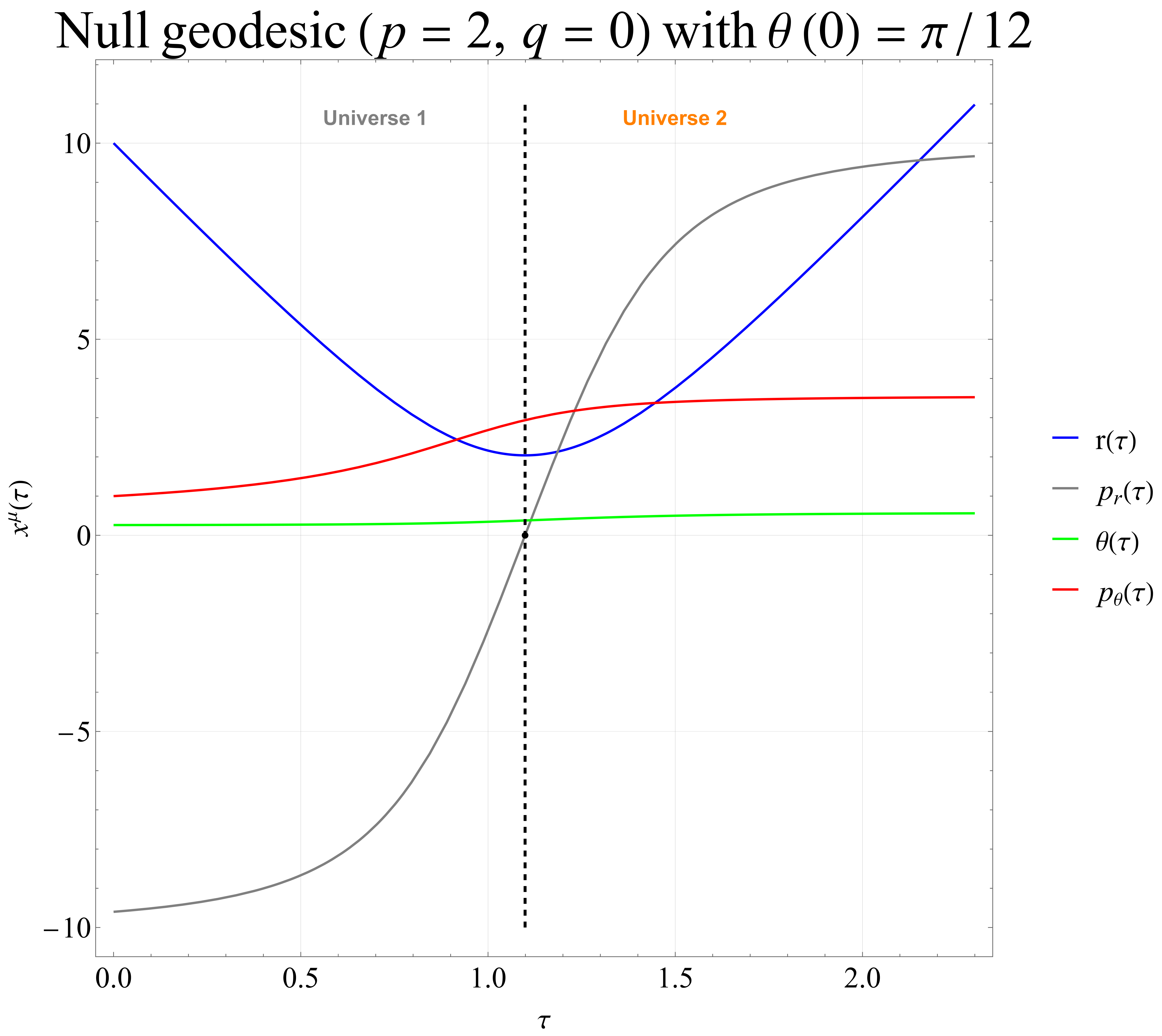}
        \subcaption{The solutions concerning the momenta and variables associated with \eqref{Solucion p2 q0}, using $J=0$, $E=10$, $r_0=2$, $p_\phi =1$, and \eqref{Condiciones iniciales p=2 pi12}. The black dotted line signifies the moment when the null geodesic touch the throat of the wormhole located in $R_G=r_0$, and this line separates universe 1 from universe 2.}
        \label{fig:GeoNulap2pi12}
    \end{minipage}
    \hfill
    \begin{minipage}{0.48\textwidth}
    \centering
    \includegraphics[width=\textwidth]{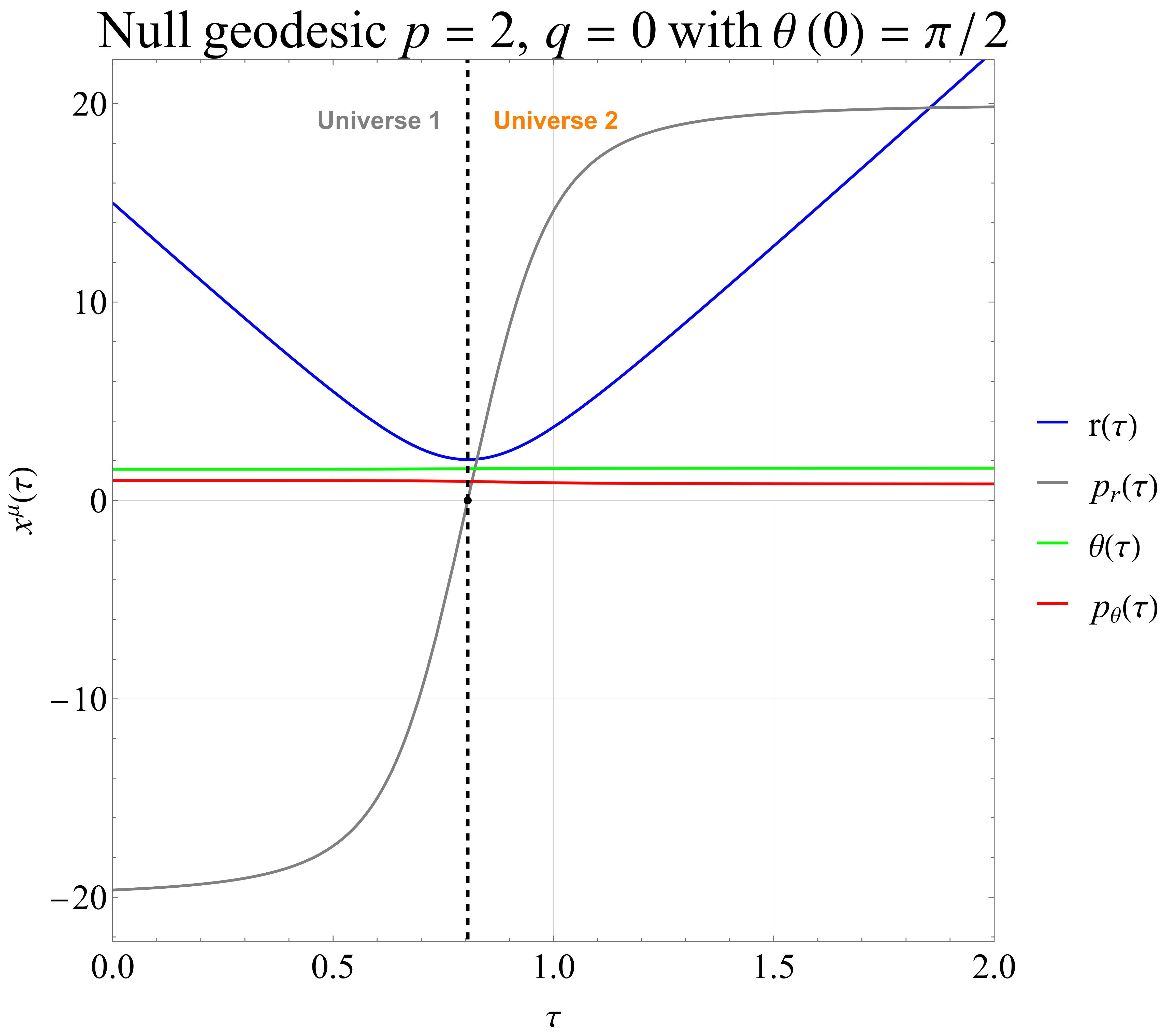}
        \subcaption{The solutions related to the momenta and variables linked to \eqref{Solucion p2 q0} are achieved with the parameters $r_0=R_G=2$, $E=20$, $J=0$, $p_\phi=10$, $r(0)=15$, and $p_\theta(0)=1$, while ensuring the condition $H(x^{\mu}(0),p_\nu(0))=0$ is met to determine $p_r(0)$. The black dotted line indicates the point at which the null geodesic reaches the throat of the wormhole at $R_G=r_0$, marking the boundary between universe 1 and universe 2.}
        \label{fig:GeoNulap2pi6}
    \end{minipage}
    \caption{}
\end{figure}

By applying the initial condition \eqref{Condiciones iniciales p=2 pi12} or \eqref{Condiciones iniciales p=2 pi2} to \eqref{Ecuaciones de Hamilton}, it is possible to construct the solution to momenta and coordinates for the initial angle $\theta=\pi/12$, see Figure \ref{fig:GeoNulap2pi12}.

It is noteworthy that the numerical solutions $r(\tau)$ maintain a uniform structure and  minimum at $R_G=r_0$, occurring at the moment $\tau_0$ depicted in figure \ref{fig:GeoNulap2pi12} as a dashed black line.

\begin{figure}[b]
    \centering
    \begin{minipage}{0.48\textwidth}
    \centering
        \includegraphics[width=\textwidth]{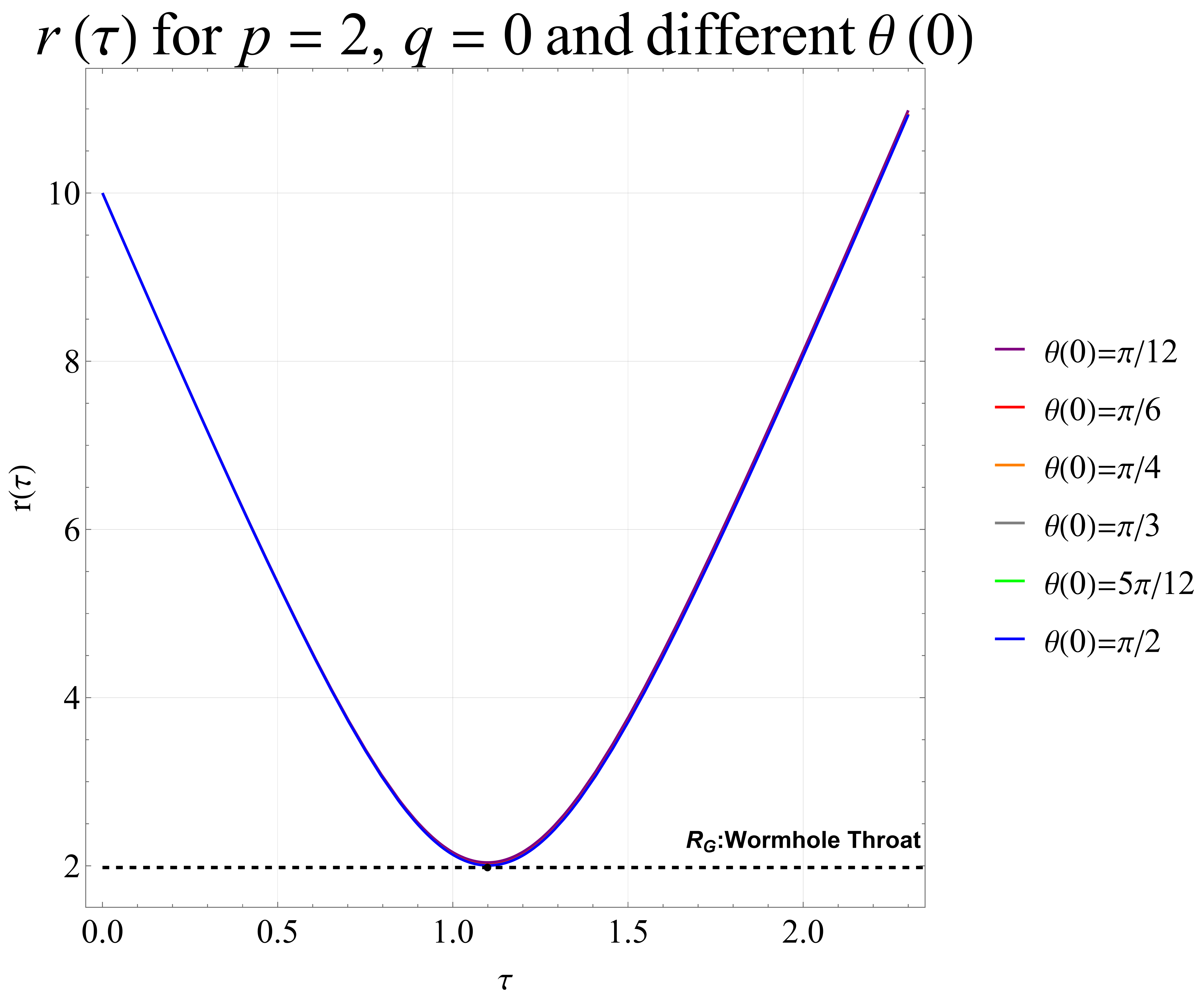}
        \subcaption{The numerical solution $r(\tau)$ for various initial $\theta$ values that allow it to reach the throat and pass through the wormhole, considering $r_0=R_G=2$, $E=10$, $J=0$, $p_\phi=1$, $r(0)=10$, and $p_\theta(0)=1$, while ensuring $H(x^{\mu}(0),p_\nu(0))=0$ to obtain $p_r(0)$.}
        \label{fig:CruceGeos1}
    \end{minipage}
    \hfill
    \begin{minipage}{0.48\textwidth}
    \centering
    \includegraphics[width=\textwidth]{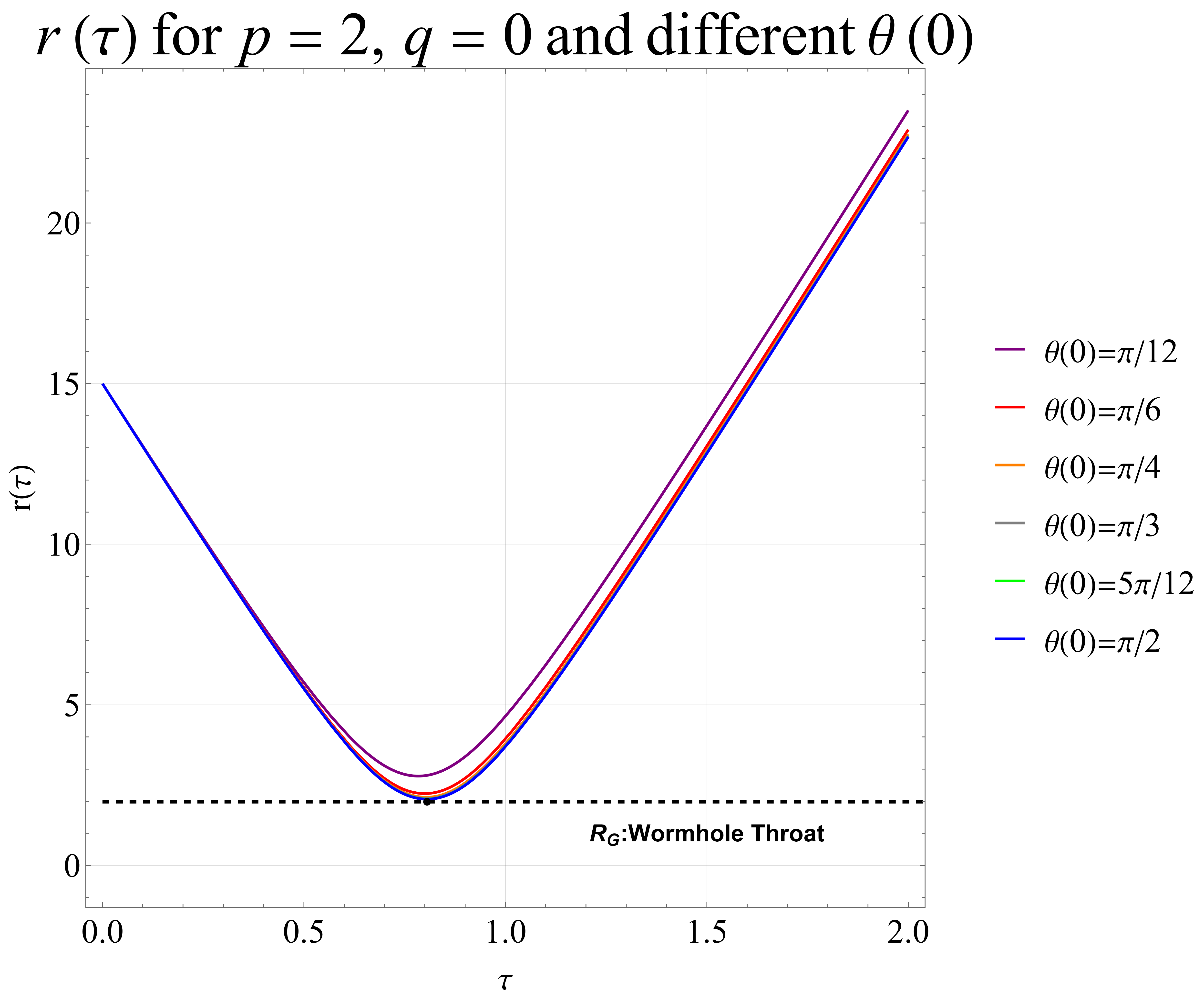}
        \subcaption{The numerical solution $r(\tau)$ for different initial $\theta$ values demonstrates that solely the geodesic with $\theta(0)=\pi/2$ is able to intersect the wormhole throat. This is achieved with parameters $r_0=R_G=2$, $E=20$, $J=0$, $p_\phi=10$, $r(0)=15$, and $p_\theta(0)=1$, while maintaining the condition $H(x^{\mu}(0),p_\nu(0))=0$ to obtain $p_r(0)$.}
        \label{fig:CruceGeos2}
    \end{minipage}
    \caption{}
\end{figure}

By employing the identical methodology utilised to derive \eqref{Condiciones iniciales p=2 pi2} and \eqref{Condiciones iniciales p=2 pi12}, but now for $\theta(0)=\{\pi/6,\pi/4,\pi/3,5\pi/12\}$, and applying the resultant conditions to obtain the numerical solution and its graphical representation, it becomes evident that for all $\theta(0)=\theta_0\in(0,\pi)$, the geodesics invariably traverse the wormhole; in other words, they touch the wormhole throat (see Figure \ref{fig:CruceGeos1}).

By employing a set of distinct parameters ($r_0=R_G=2$, $E=20$, $J=0$, $p_\phi=10$, $r(0)=15$, and $p_\theta(0)=1$), we are able to generate Figures \ref{fig:CruceGeos2} and \ref{fig:GeoNulap2pi6}. The first figure depicts the functions $r(\tau)$ for various initial values of $\theta(0)$, with the dashed black line denoting the wormhole throat. The second figure illustrates the momenta and coordinates for the geodesic with $\theta(0)=\pi/2$, which uniquely possesses the capability to traverse the wormhole. The dashed black line once again signifies the moment at which the geodesic crosses the wormhole.

\begin{figure*}
    \begin{minipage}{0.65\textwidth}
        \centering
        \begin{subfigure}{\textwidth}
            \includegraphics[width=\textwidth]{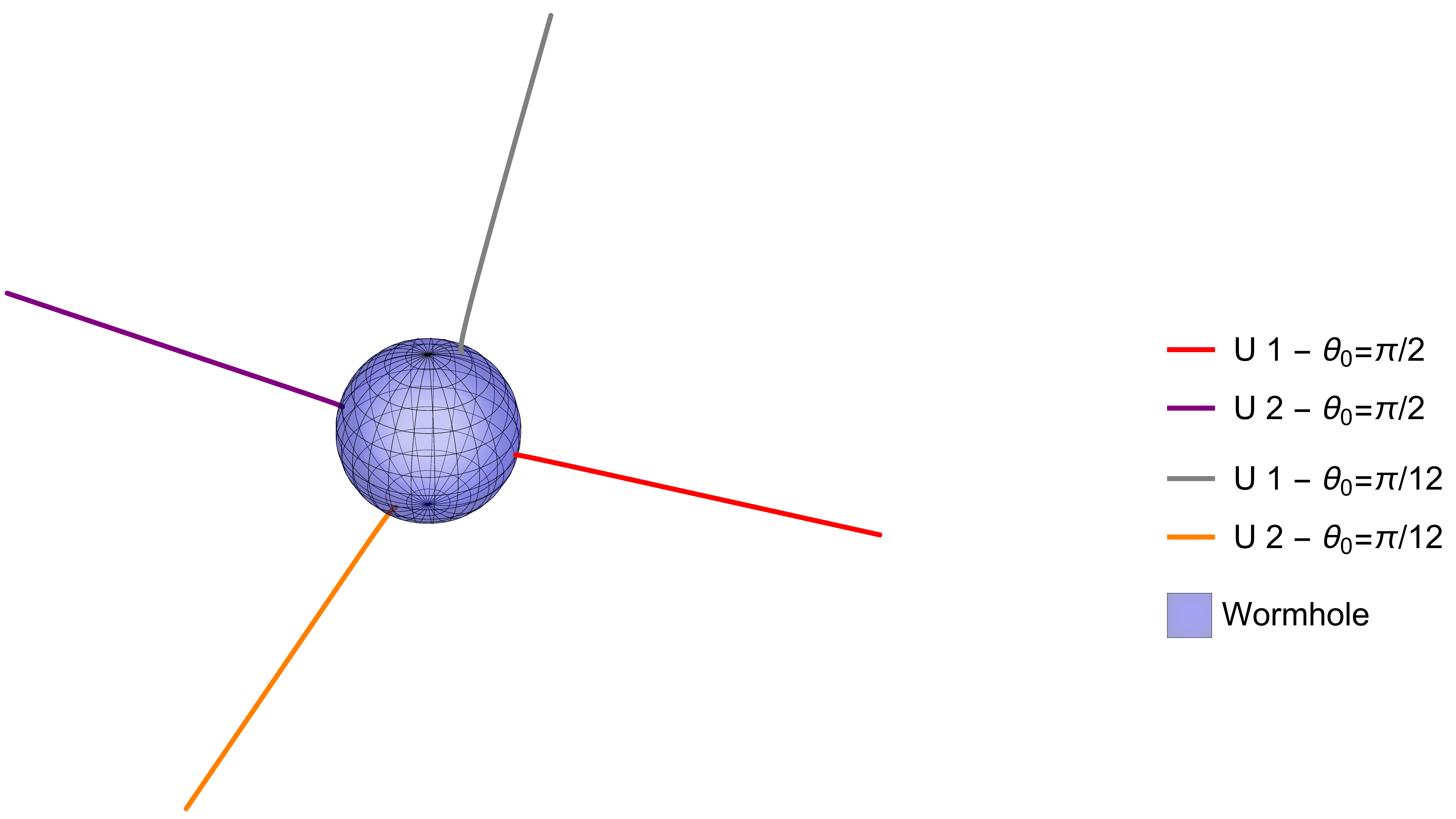}
            \caption{The duo of null geodesics progress through the wormhole, making contact with the throat (depicted as a sky blue sphere) and continuing their trajectory into the alternate universe. The graphing is based on the parameters from Figure \ref{fig:GeoNulap2pi12}.}
            \label{fig:Geo3Dp2}
        \end{subfigure}
    \end{minipage}%
    \hfill
    \begin{minipage}{0.65\textwidth}
        \centering
        \begin{subfigure}{\textwidth}
            \includegraphics[width=\textwidth]{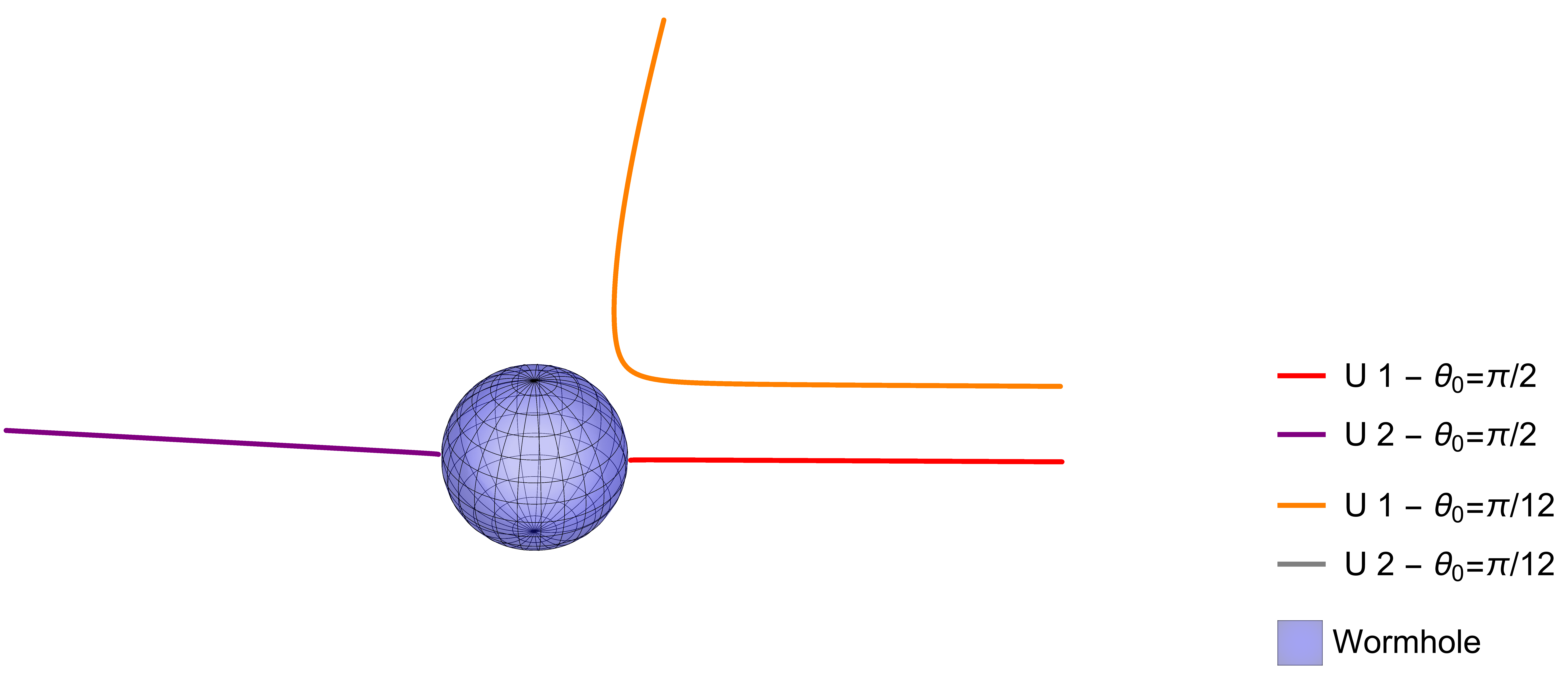}
            \caption{Only the null geodesics in the equatorial plane progress through the wormhole, making contact with the throat (depicted as a sky blue sphere) and continuing their trajectory into the alternate universe. The nearby null geodesic close to the polar axis is unable to pass through the wormhole and remains within the same universe. The graphing is based on the parameters from Figure \ref{fig:GeoNulap2pi6}.}
            \label{fig:Geo3Dp2Muchas}
        \end{subfigure}
    \end{minipage}%
    
    \caption{Null geodesics are depicted in three dimensions alongside the throat's shape, highlighted in sky blue and overlaid with a mesh grid to enhance visualization. In the label of each subfigure, the letter U signifies \textit{Universe}.}
    \label{fig:Geo3D}
\end{figure*}

In Figure \ref{fig:Geo3D} and \ref{fig:Geo3DCaso2}, we provide an in-depth illustration of three-dimensional geodesics represented with Cartesian coordinates with $\phi (\tau)=g^{\phi \phi} p_\phi$:

\begin{align*}
    X&=r(\tau) \cos{(\phi (\tau))} \sin{(\theta(\tau))},\\
    Y&=r(\tau) \sin{(\phi (\tau))} \sin{(\theta(\tau))},\\
    Z&=r(\tau) \cos{(\phi (\tau))}.
\end{align*}

In \ref{fig:Geo3Dp2}, we illustrate two distinct geodesics, given by initial conditions \eqref{Condiciones iniciales p=2 pi12} and \eqref{Condiciones iniciales p=2 pi2}. To enhance clarity, we employ four colors—two for each geodesic to illustrate when the corresponding curves transition to another universe. Similarly, we depict the structure of the wormhole throat, corresponding to a $\mathbb{S}^2$ sphere with a radius of $R_G=r_0$. The internal region $\mathbb{B}_{r_0}=\{\Vec{r}\in \mathbb{R} | \parallel \Vec{r} \parallel <R_G \}$ signifies the domain wherein closed timelike curves (CTCs) manifest, however, all geodesics initially intersect the wormhole throat prior to the transgression of causality. On the other hand, Figure \ref{fig:Geo3Dp2Muchas} illustrates the second example of conditions, where the only geodesic that permits traversal through the wormhole is located proximal to the equatorial plane.

\subsubsection{\texorpdfstring{Plotting for $q=6\sqrt{3}$}{Values q=6R3}}

Repeat the procedure described in the previous subsection again, but this time with the following parameters: $ p=2,q=6\sqrt{3},E=1,J=0,p_\phi=1,R_G=3.47197$ and the initial conditions $r(0)=5,p_\theta(0)=1$. This allows us to plot the null geodesics for these new solutions, which exhibit unusual oscillation behaviour.

\begin{figure}[b]
    \centering
    \begin{minipage}{0.48\textwidth}
    \centering
        \includegraphics[width=\textwidth]{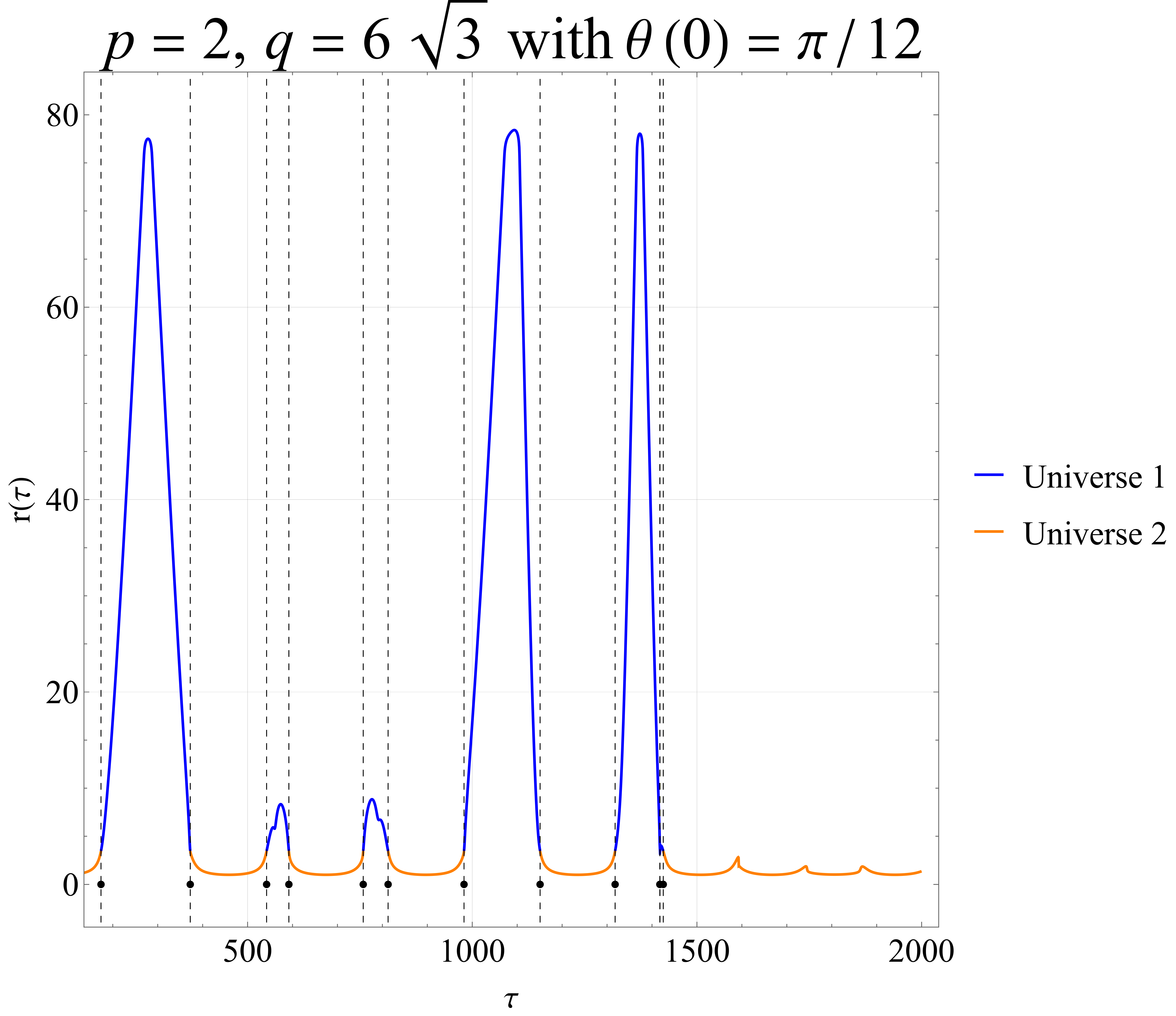}
        \subcaption{Radial function close to the polar axis that relates to the solution $p=2$, $q=6\sqrt{3}$, where $p_r(0)=-18.7628$. The dashed black lines indicate the proper time as the geodesic crosses the wormhole. The blue line represents one universe, while the orange line signifies another. The throat size is given as $R_G=3.47197$, and the ergo-region radius is $r_0=1<R_G$.}
        \label{fig:FuncionRadialPiDoceavos}
    \end{minipage}
    \hfill
    \begin{minipage}{0.48\textwidth}
    \centering
    \includegraphics[width=\textwidth]{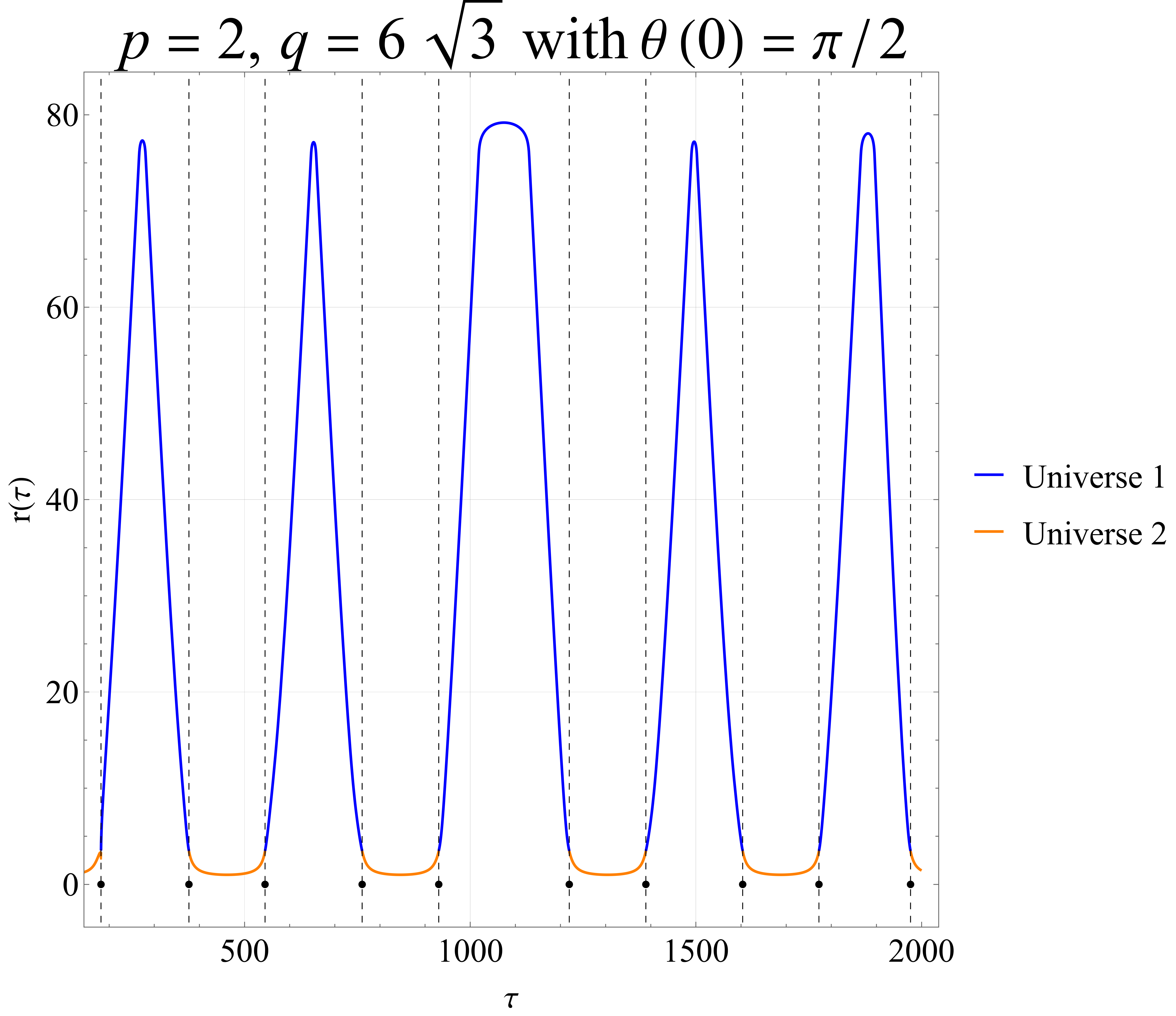}
        \subcaption{Radial function graph near the equatorial plane associated with the solution $p=2$, $q=6\sqrt{3}$, where $p_r(0)=-4.25258$. The dashed black lines show the proper time as the geodesic traverses the wormhole. The blue line denotes one universe, while the orange line represents another. The throat size is indicated as $R_G=3.47197$, and the ergo-region radius is $r_0=1<R_G$.}
        \label{fig:FuncionRadialPiMedios}
    \end{minipage}
    \caption{}
\end{figure}

\begin{figure*}
    \begin{minipage}{0.65\textwidth}
        \centering
        \begin{subfigure}{\textwidth}
            \includegraphics[width=\textwidth]{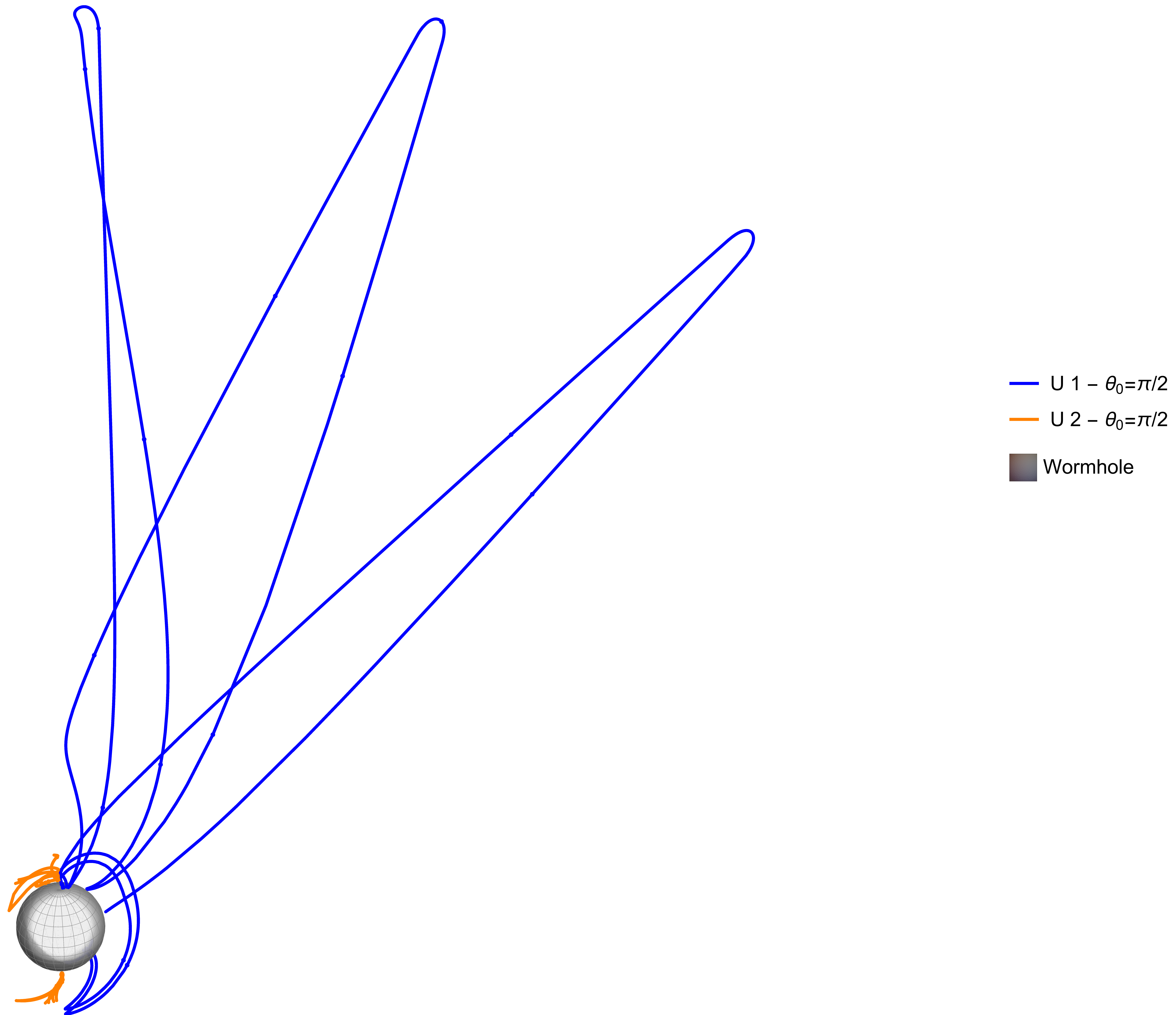}
            \caption{The graph that corresponds to the solution $p=2,q=6\sqrt{3}$ utilizes the parameters from Figure \ref{fig:FuncionRadialPiDoceavos}. This graph illustrates the geodesics with $\theta(0)=\pi/12$ when the particle tends to stays within a single universe.}
            \label{fig:Geo3Dp2q6r3EjePolar}
        \end{subfigure}
    \end{minipage}%
    \hfill
    \begin{minipage}{0.65\textwidth}
        \centering
        \begin{subfigure}{\textwidth}
            \includegraphics[width=\textwidth]{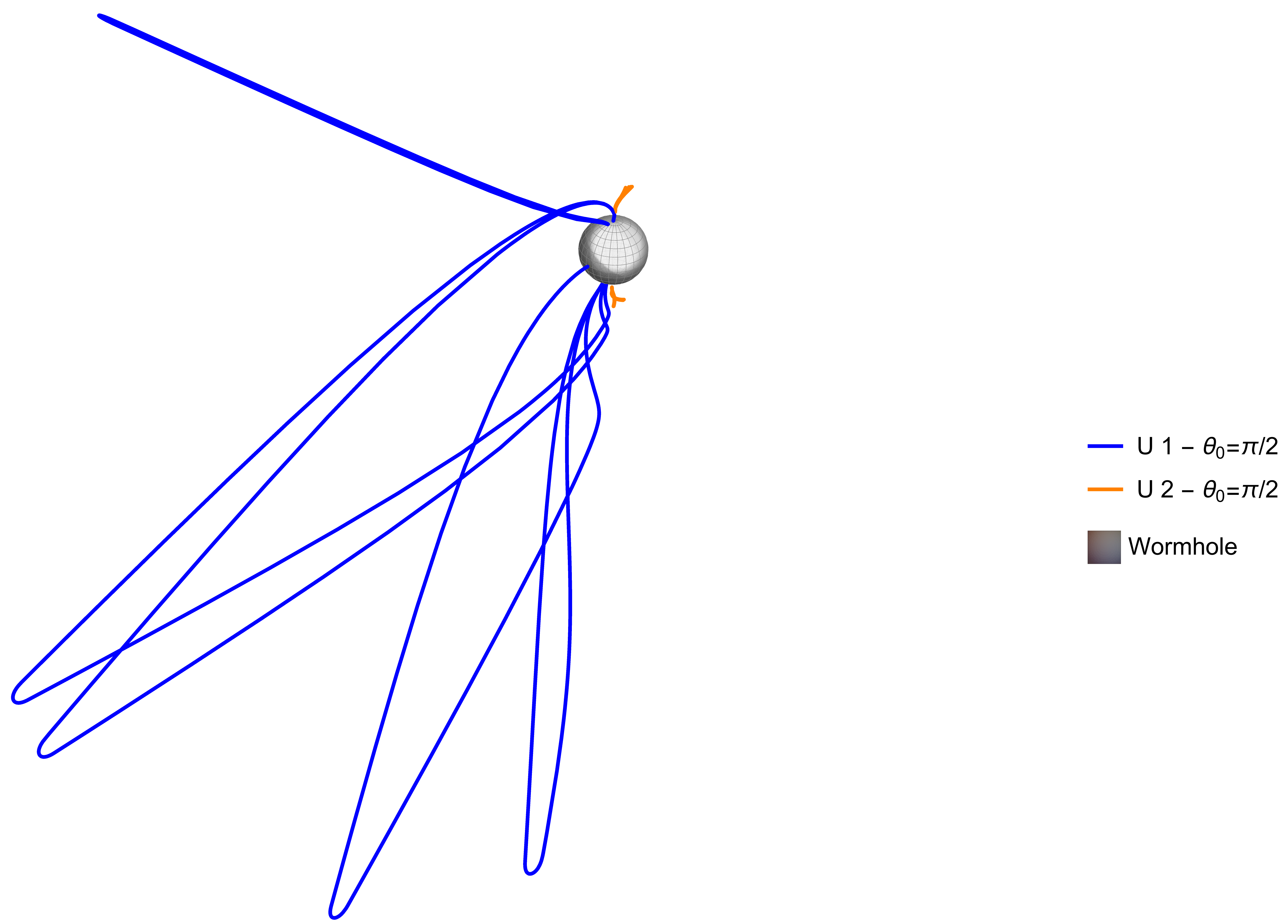}
            \caption{The graph that corresponds to the solution $p=2,q=6\sqrt{3}$ employs the parameters from Figure \ref{fig:FuncionRadialPiMedios}. This graph depicts the geodesics with $\theta(0)=\pi/2$, transitioning between two universes.}
            \label{fig:Geo3Dp2q6r3PlanoEcuatorial}
        \end{subfigure}
    \end{minipage}%
    
    \caption{Null geodesics are schematically depicted in three dimensions alongside the throat's shape, highlighted in sky blue and overlaid with a mesh grid to enhance visualization. In the label of each subfigure, the letter U signifies \textit{Universe}.}
    \label{fig:Geo3DCaso2}
\end{figure*}

Figure \ref{fig:FuncionRadialPiMedios} illustrates the radial function related to a geodesic directed close to the equatorial plane. We can see that its behaviour is oscillatory as it shifts from one universe to another. Importantly, it does not exceed a radius of $r_{U2}=R_G+r_0$. This suggests that the causal universe within universe 2 might exclusively be a sphere with a radius of $r_{U2}$. Universe 1 corresponds to the blue curves, while universe 2 is linked to the orange curves. Meanwhile, the dashed black lines indicate when photons travel between the two universes. These geodesics, positioned near the equatorial plane in Cartesian coordinates, are also shown in Figure \ref{fig:Geo3Dp2q6r3PlanoEcuatorial}.

On the other hand, Figure \ref{fig:FuncionRadialPiDoceavos} illustrates a scenario near the polar axis, as opposed to being close to the equatorial plane. Here, the oscillation pattern changes because the photon covers more distance in universe 1 (indicated in blue) during certain intervals, while in others, it does not. Moreover, beyond a specific proper time, the geodesic seems to remain in universe 2 (highlighted in orange), showing a different behaviour from the previous case. Additionally, the causal universe 2 appears to have a size of $r_{U2}$. Refer to Figure \ref{fig:Geo3Dp2q6r3EjePolar} for the relevant 3D Cartesian plot.

\section{Conclusions}
\label{section: conclusions}

In this work, we used the flat subspaces method to build solutions to the 5-dimensional EFE in vacuum.
These solutions depend on two constant parameters $p$ and $q$.
We only studied the case when $p$ is a natural number; however, the solutions are also valid for any value of $p$.

The Kreschman invariant of these solutions depends on the constant parameter $l$.
For $l = 0$ and $l \leq -\frac{1}{2}$, the solutions are regulars.
The metrics $\hat{g}_p$ are regular when $\vert q \vert = \sqrt{\frac{ p^2 - 4 }{3}}$ for $p \geq 2$, or if $\vert q \vert \leq \sqrt{ \frac{p^2}{3} - 2 }$ for $p \geq 3$.

We found that the metrics $\hat{g}_p$ with even $p$ describe wormholes.
One universe is defined by $r > R_G$ and the other by $r < -R_G$.
Both universes have $\frac{p}{2}$ ergoregions.
The larger ergoregion is given by $r = r_e$ for one universe, and for the other it is $r = -r_e$.

We determined that null geodesics are well defined for $r \geq r_e$.
However, there are points inside the larger ergoregion where the null geodesics are not defined, because the motion equation for $r$ gives $\dot{r}^2 < 0$.

We defined the function $F ( x, m )$, which allowed us to find that $R_G = r_e$ for $l \leq 0$.
For the case $l > 0$, $R_G = r_e$ holds if $q$ satisfies at least one of the following inequalities: $q < 0$ for all $p \geq 2$, $q \leq \frac{1 + \sqrt{2}}{3} p$ for all $p \geq 4$ and $q \leq \frac{1}{3}$ when $p = 2$.

By examining the geodesics, it was deduced that in the particular instance where $q=0$, the solutions feature a throat at $R_G=r_0=r_e$, indicating these wormholes are regular and uphold causality in alignment with the Wormhole Cosmic Censorship Conjecture \cite{axioms14110831}. Nonetheless, in the scenario where $q=6\sqrt{3}$, the geodesics display an unusual oscillating pattern. When a photon is directed near the polar axis, its trajectory is non-periodic, resulting in the geodesic remaining in universe two after a certain period. On the other hand, when directed near the equatorial plane, the geodesic continues to oscillate between one universe and the other. Crucially, the geodesic cannot exit the $\mathbb{S}^2$ sphere with radius $r_{U2}=R_G+r_0$, as doing so would breach causality, in simple terms.

Most significantly, we were able to mathematically demonstrate and provide a numerical example showing that the throats of these wormholes with $q \neq 0$ adhere to the WCCC under specific conditions.
The WCCC is satisfied if $l \leq 0$ and for $l > 0$ the inequalities $q < 0$ for all $p \geq 2$, $q \leq \frac{1}{3}$ for $p  = 2$ and $q \leq \frac{1 + \sqrt{2}}{3} p$ for all $p \geq 4$ are fulfilled, indicating that the throat conceals the singularities present at $r=0$ and $\theta=\pi/2$ and dress the CTC region with $r_e<R_G$.


\begin{appendices}


\section{Method Explanation}
\label{Apendice: Method Explanation}

In \cite{Sarmiento-Alvarado2025}, the authors introduce a new method to find exact solutions to vacuum Einstein field equations (EFE) in higher dimensions.
This method considers a $n + 2$-dimensional spacetime endowed with a metric $\hat{g}$ that admits $n$ commutative Killing vectors.
Under this assumption, there exists a system of coordinates such that the metric $\hat{g}$ has the form
\begin{equation}
\label{metric}
    \hat{g}
    = f ( d\alpha^2
    + d\zeta^2 )
    + g_{\mu \nu} dx^\mu dx^\nu
\end{equation}
for all $\mu, \nu \in \{ 3, \ldots, n + 2 \}$, where the metric components $f$ and $g_{\mu \nu}$ depend on variables $\alpha$ and $\zeta$.
Then, the vacuum EFE ($R_{A B}=0$ for all $A, B \in \{ 1, \ldots, n + 2 \}$) are
\begin{subequations}
\begin{align}
\label{chiral eq g}
&    ( \alpha g_{, w} g ^{-1} )_{, \bar w} + ( \alpha g_{, \bar w} g ^{-1} )_{, w} = 0
,
\\\label{SL invariant field eq f}
&
    ( \ln f \alpha ^{1-1/n} )_{, W} = \frac{\alpha}{2} \operatorname{tr} ( g_{, _W} g^{-1} )^2
    \text{ for } W = \{ w, \bar w \},
\end{align}
\end{subequations}
where $g$ is defined as $( g )_{\mu \nu} = -\alpha^{-2/n} g_{\mu \nu}$, $\alpha = \sqrt{ -\det g_{\mu \nu}}$ and $w = \alpha + i \zeta$.
Note that $g$ is a symmetry matrix and belongs to the Lie group $SL ( n, \mathbb{R} )$.

In order to solve the chiral equation \eqref{chiral eq g}, they assume that $g$ depends on a set of parameters $\xi^a = \xi^a ( w, \bar{w} )$, that is, $g = g ( \xi^a ( w, \bar{w} ) )$, that satisfy the generalized Laplace equation:
\begin{equation}
\label{gen Laplace eq}
    \left( \alpha \xi^a _{, w} \right)_{, \bar w}
    + \left( \alpha \xi^a _{, \bar w} \right)_{, w}
    = 0 .
\end{equation}
Then, the chiral equation (\ref{chiral eq g}) becomes
\begin{equation}
\label{Killing eq A Riemannian space}
    A_{a , b} + A_{b , a} = 0
    \text{ for all } a, b \in \{ 1, \ldots, r \} ,
\end{equation}
where $_{, a}$ denote the partial derivative with respect to $\xi^a$, $1 \leq r < n$ and the matrices $A_a$ are defined as
\begin{equation}
\label{def matrices A}
    A_a = g_{, a} g^{-1} .
\end{equation}
This method assumes that the matrices $A_a$ are constants, which implies that the matrices $A_a$ commute with each other, as a consequence of $A_{a, b} = -\frac{1}{2} [A_a, A_b] = 0$.

Since $\det g = (-1)^{n + 1}$, then $\operatorname{tr} A_a = 0$, so that the matrices $A_a$ belong to the Lie algebra $\mathfrak{sl} ( n, \mathbb{R} )$.
Under transformations $g \to C g C ^T$, with $C \in SL ( n, \mathbb{R} )$ constant, the chiral equation (\ref{chiral eq g}) remains unchanged.
Consequently, the matrices $A_a$ transform into $A_a \to C A_a C^{-1}$.
Thus, the set of matrices $A_a$ is divided into equivalence classes.
In \cite{Sarmiento-Alvarado2023}, the authors classify the equivalence classes of $\mathfrak{sl} ( n, \mathbb{R} )$ into five different types according to their eigenvalues.

In \cite{Sarmiento-Alvarado2025}, the authors propose two techniques to find a set $\{ A_a \}$ of pairwise commuting matrices.
The first is based on the centralizer of a subset $\mathscr{A} \subset \mathbf{M}_n$, which is defined as
\begin{equation}
    \mathcal{C} (\mathscr{A})
    = \{ M \in \mathbf{Sym}_n : A M = M A \text{ for all } A \in \mathscr{A}\} .
\end{equation}
In this technique, the matrix $A_1$ is representative of some equivalence class of $\mathfrak{sl} ( n, \mathbb{R} )$.
Then, the matrix $A_i$ is chosen from $\mathcal{C} ( A_1, \ldots, A_{i - 1} )$ for all $i \in \{ 2, \ldots, r \}$.

The other technique uses the fact that there exists a commutative algebra $\mathfrak{A}$ of dimension $n$ contained in the centralizer of an equivalence class of $\mathfrak{sl} ( n, \mathbb{R} )$.
Thus, the set $\{ A_a \}$ is selected from $\mathfrak{A}$.

Due to Eq. \eqref{def matrices A}, Eq. \eqref{SL invariant field eq f} is modified as
\begin{equation}
    ( \ln f \alpha ^{1-1/n} )_{, W}
    = \frac{\alpha}{2}  \operatorname{tr} ( A_a A_b ) \xi^a_{, W} \xi^b_{, W} .
\end{equation}
Solving it for a set $\{ \xi^a ( w, \bar{w} ) \}$ of solutions of Eq. \eqref{gen Laplace eq}, we find the function $f$.
The generalized Laplace equation \eqref{gen Laplace eq} is solved in \cite{Matos:2010pcd,Sarmiento-Alvarado2025}.

Once we know a set $\{ A_a \}$ of pairwise commuting matrices and a set $\{ \xi^a ( w, \bar{w} ) \}$ of solutions to the generalized Laplace equation \eqref{gen Laplace eq}, we compute the solution to the chiral equation \eqref{chiral eq g} given by
\begin{equation}
\label{chiral eq sol}
    g ( w, \bar{w} ) = \exp\left( \xi^a ( w, \bar{w} )  A_a \right) g_0 ,
\end{equation}
where $g_0$ is a constant matrix in $\mathcal{I} (\{ A_a \})$.
The set $\mathcal{I} (\mathscr{A})$ is defined as
\begin{equation}
    \mathcal{I} (\mathscr{A})
    = \{ X \in \mathbf{Sym}_n : A X = X A^T \text{ for all } A \in \mathscr{A}\} .
\end{equation}
In \cite{Sarmiento-Alvarado2023}, the authors compute the solutions for $r = 1$, that is, $g ( w, \bar{w} ) = \exp\left( \xi ( w, \bar{w} )  A \right) g_0$ for each of the five types of equivalence classes of $\mathfrak{sl} ( n, \mathbb{R} )$, while the solutions involving commutative algebras $\mathfrak{A}$ are determined in \cite{Sarmiento-Alvarado2025}.




\section{Complete function}
\label{Apendice: Complete function}

\begin{widetext}
\begin{align}\label{eq:F0}
    \mathcal{F}_0
&   = -96 r^4 (p^2-3 q^2 )+96 q r_0 r^3 (3 p^2-5 q^2 )+6 q r_0^3 r  (3 p^2-5 q^2) (\cos (2 \theta ) (p^2-3 q^2+4)-p^2+3  (q^2+4 ) ) \notag\\
    &-12 r_0^2 r^2  (\cos (2 \theta )  (p^2-3 q^2)^2-2 (p^4-2 p^2 (5 q^2+2)+q^2 (13 q^2+12))) \notag\\
    &+r_0^4 (-p^6+p^4 (9 q^2+14)-p^2 (27 q^4+132 q^2+16)+\cos (2 \theta ) (p^6-p^4 (9 q^2+2)\notag \\
    &+p^2 (27 q^4-36 q^2+16)-3 q^2 (9 q^4-10 q^2+16))+3 q^2 (9 q^4+58 q^2+16)) .
\end{align}
\end{widetext}


\section{\texorpdfstring{Solution $p=2$ and $q=0$}{Values p=2 q=0}} 
\label{Apendice:Solucion p2 q0}

By setting the parameters $p=2, q=0, \mu=-1, \nu=0$ in the general metric \eqref{Solucion p q generales}, we can derive:
\begin{equation}\label{Solucion p2 q0}
\begin{aligned}
    &\hat{g}_2 =
   \frac{
        dr^2
        + \Delta_r ( d\theta^2
    + \sin^2 \theta d\phi^2 )
    }{\Xi^2} 
\\& + \frac{ r^2 - r_0^2 }{ r^2 + r_0^2 } \Big( 
        d\psi^2 -dt^2
    )
    -\frac{ 4 r_0 r}{ r^2 + r_0^2 } dt d\psi .
\end{aligned}
\end{equation}

\end{appendices}

\bibliographystyle{elsarticle-harv} 
\bibliography{Bibliografia}

\end{document}